\documentclass[%
reprint,
superscriptaddress,
nofootinbib,
 amsmath,amssymb,
 aps,
]{revtex4-1}

\usepackage{placeins}
\usepackage{graphicx}% Include figure files
\usepackage{dcolumn}% Align table columns on decimal point
\usepackage{bm}% bold math
\usepackage{chemformula}% chemie symbols
\usepackage{textcomp}
\usepackage{upgreek}
\usepackage{xcolor}
\usepackage[normalem]{ulem}
\usepackage{braket}

\usepackage[pagebackref=false]{hyperref}
\usepackage{upgreek}

\newcommand{\lr} {\lambda_\text{R}}
\newcommand{\lvz} {\lambda_\text{VZ}}

\begin{document}

\title{Weak localization as probe of spin-orbit-induced spin-split \\ bands in bilayer graphene proximity coupled to \ch{WSe2}}

\author{E.~Icking}
\affiliation{JARA-FIT and 2nd Institute of Physics, RWTH Aachen University, 52074 Aachen, Germany,~EU}%
\affiliation{Peter Gr\"unberg Institute  (PGI-9), Forschungszentrum J\"ulich, 52425 J\"ulich,~Germany,~EU}

\author{F.~Wörtche}
\affiliation{JARA-FIT and 2nd Institute of Physics, RWTH Aachen University, 52074 Aachen, Germany,~EU}%

\author{A.W.~Cummings}
\affiliation{Catalan Institute of Nanoscience and Nanotechnology (ICN2), CSIC and BIST, Campus UAB, Bellaterra, 08193 Barcelona, Spain,~EU}%

\author{A.~Wörtche}
\affiliation{JARA-FIT and 2nd Institute of Physics, RWTH Aachen University, 52074 Aachen, Germany,~EU}%

\author{K.~Watanabe}
\affiliation{Research Center for Functional Materials, 
National Institute for Materials Science, 1-1 Namiki, Tsukuba 305-0044, Japan
}
\author{T.~Taniguchi}
\affiliation{ 
International Center for Materials Nanoarchitectonics, 
National Institute for Materials Science,  1-1 Namiki, Tsukuba 305-0044, Japan
}%
\author{C.~Volk}
\affiliation{JARA-FIT and 2nd Institute of Physics, RWTH Aachen University, 52074 Aachen, Germany,~EU}%
\affiliation{Peter Gr\"unberg Institute  (PGI-9), Forschungszentrum J\"ulich, 52425 J\"ulich,~Germany,~EU}

\author{B. Beschoten}
\affiliation{JARA-FIT and 2nd Institute of Physics, RWTH Aachen University, 52074 Aachen, Germany,~EU}%
\author{C. Stampfer}
\affiliation{JARA-FIT and 2nd Institute of Physics, RWTH Aachen University, 52074 Aachen, Germany,~EU}%
\affiliation{Peter Gr\"unberg Institute  (PGI-9), Forschungszentrum J\"ulich, 52425 J\"ulich,~Germany,~EU}%

\date{\today}% It is always \today, today,
             %  but any date may be explicitly specified


\begin{abstract} 
Proximity coupling of bilayer graphene (BLG) to transition metal dichalcogenides (TMDs) offers a promising route to engineer gate-tunable spin–orbit coupling (SOC) while preserving BLG's exceptional electronic properties. This tunability arises from the layer-asymmetric electronic structure of gapped BLG, where SOC acts predominantly on the layer in contact with the TMD. Here, we present a high-quality BLG/\ch{WSe2} device with a proximity-induced SOC gap and excellent electrostatic control. Operating in a quasi-ballistic regime, our double-gated heterostructures allow to form gate-defined p–n–p cavities and show clear weak anti-localization (WAL) features consistent with Rashba-type SOC. At lower hole densities, a transition to weak localization (WL) is observed, signaling transport through a single spin-split valence band. These findings - in agreement with calculations - provide direct spectroscopic evidence of proximity-induced spin-split band in BLG and underscore the potential of BLG/TMD heterostructures for spintronics and spin-based quantum technologies.

\end{abstract}

\maketitle

\begin{figure}[hbt]
	\centering
\includegraphics[draft=false,keepaspectratio=true,clip,width=01.0\linewidth]{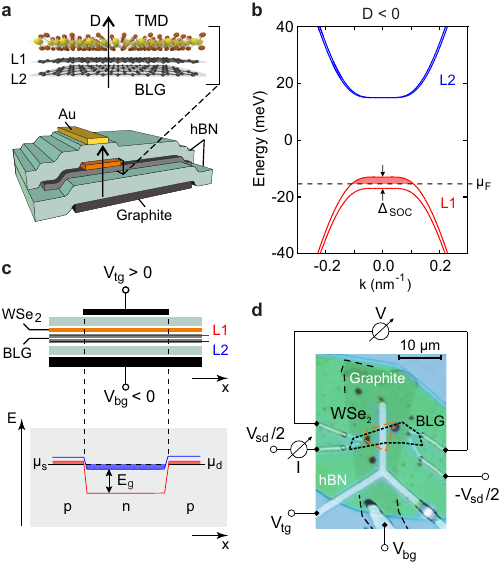}
\caption[Fig01]{
    \textbf{(a)} Schematics of the dual-gated BLG/TMD  device.
    \textbf{(b)} Calculated band structure of BLG proximity-coupled to \ch{WSe2} for a negative displacement field $D/\varepsilon_0$. In this case the valance band is spin-split by $\Delta_{\text{SOC}}$.
    \textbf{c} Schematic cross section of the device highlighting the proximal layer (L1) and the second graphene layer  of the BLG (L2).
    The top gate allows the tuning of the Fermi level independently of the applied $D$-field. For $V_\mathrm{bg}<0$, there will be hole (p)-doping of the BLG regions that are not covered by the top gate. Positive top gate voltages ($V_\mathrm{tg}>0$) can create an n-doped region underneath the top gate, leading to the formation of lateral p-n junctions at the boundaries of the top gate (depicted in the lower panel). 
    \textbf{d} False color image of the sample highlighting the different layered materials and electrical contacts of the device including the measurement scheme.
}
\label{f1}
\end{figure}

Combining graphene with transition metal dichalcogenides (TMDs) allows one to enhance the weak intrinsic spin-orbit coupling (SOC) strength of graphene while maintaining its high carrier mobility~\cite{Gmitra2015oct, Gmitra2016apr, Wang2015nov, Wang2016oct, Yang2017jul, Voelkl2017Sep, Zihlmann2018Feb, Wakamura2018Mar, Banszerus2019Sep,Fulop2021Sep,Tiwari2022Oct,Bisswanger2025May}. 
This enhanced SOC enables potential applications such as spin valves~\cite{Garcia2018, Ahn2020}, spin polarizers~\cite{Cummings2017Nov, Ingla2021Jul}, or spin logic gates~\cite{Han2014Oct, Hu2020Dec}.
However, the limited tunability of the proximity-induced SOC in graphene/TMD heterobilayers makes it hard to tune such spin-based devices and remains a key challenge for the realization of graphene-based spin transistors~\cite{JFabianOct2017, Levitov2017}. In contrast, double-gated Bernal-stacked bilayer graphene (BLG) proximitized to a TMD is expected to exhibit gate-tunable SOC over a wide range~\cite{Levitov2017, JFabianOct2017}.

\begin{figure*}[hbt]
	\centering
\includegraphics[draft=false,keepaspectratio=true,clip,width=0.99\linewidth]{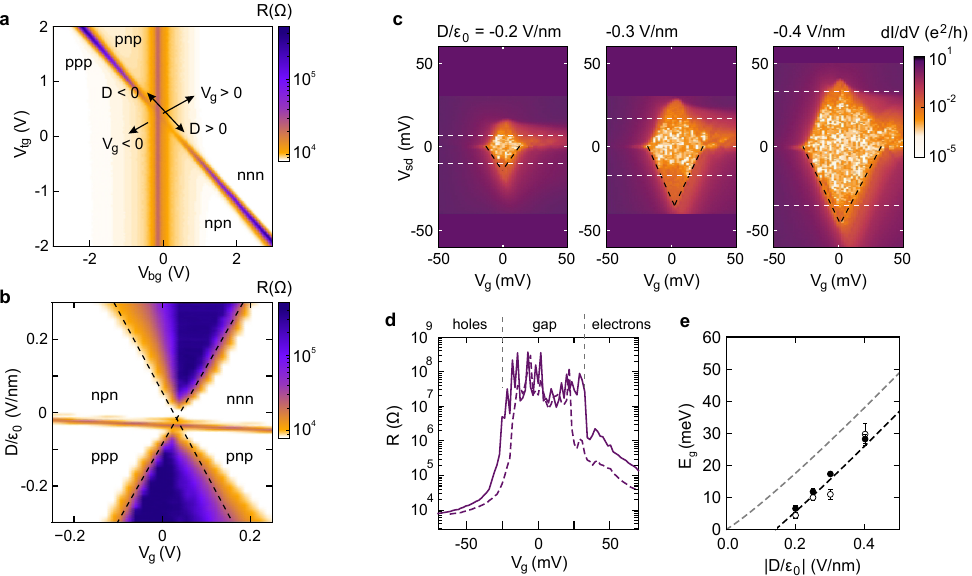}
\caption[Fig02]{
    \textbf{a} Two-terminal resistance of BLG/WSe$_2$ device as a function of top gate and bottom gate voltages ($V_\mathrm{tg,bg}$). 
    \textbf{b} Resistance as a function of the effective gate voltage $V_\mathrm{g}$ and of the electric displacement field $D/\varepsilon_0$ in the double-gated region. The measurements were performed at $T=60$\,mK, applying a small source-drain voltage of $V_\mathrm{sd}=1$~mV.
    \textbf{c} Finite bias spectroscopy measurements of BLG/\ch{WSe2} device showing the differential conductance as a function of $V_\mathrm{sd}$ and $V_\mathrm{g}$ for $D/\varepsilon_0 = -0.25\,$V/nm, $-0.3$\,V/nm and $-0.4$\,V/nm. The outline of the diamonds is highlighted by black dashed lines (assuming a threshold value of $\text{d}I/\text{d}V_\text{th}=10^{-3}e^2/h$). The white dashed lines indicate the maximal extension of the diamond along the $V_\text{sd}$ axis.
    \textbf{d} Line cuts through finite bias spectroscopy measurements at $V_\mathrm{sd} = 1.5$\,mV for displacement fields $D/\varepsilon_0 = -0.25\,$V/nm (dashed purple) and $D/\varepsilon_0 = -0.4\,$V/nm (solid purple) of the BLG/\ch{WSe2} device.
    \textbf{e} Band gap energies $E_\mathrm{g}$ of BLG extracted from finite bias spectroscopy for positive (filled circles) and negative (open circles) displacement fields (indicated with white dashed lines in panel a). The grey line shows a theoretical prediction of $E_\mathrm{g}$ according to Ref.~\cite{McCann_2013}. For the black line, an offset of 12\,meV was subtracted.
}
\label{f2}
\end{figure*}

\begin{figure*}[hbt]
	\centering
\includegraphics[draft=false,keepaspectratio=true,clip,width=1\linewidth]{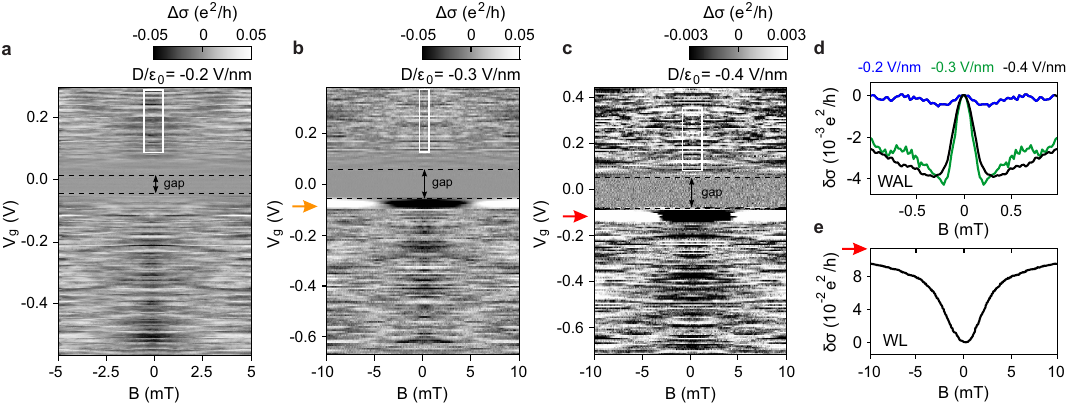}
\caption[Fig03]{
    Conductance as a function of $V_\mathrm{g}$ and $B$ for different values of the displacement field: \textbf{a} $D/\varepsilon_0 = -0.2\,$V/nm, \textbf{b} $D/\varepsilon_0 = -0.3\,$V/nm, and \textbf{c} $D/\varepsilon_0 = -0.4\,$V/nm. 
    For better visualization, we plot the conductance 
$\Delta \sigma = \sigma(B, V_\mathrm{g}=\text{const.}) - \langle \sigma(B, V_\mathrm{g}=\text{const.}) \rangle$ for each value of $V_\mathrm{g}$, where we subtract the average over $B$ of the measured conductance. 
\textbf{d} WAL peak for different values of $D/\varepsilon_0$. Here we plot $\delta \sigma = \Delta \sigma-\Delta \sigma(B=0)$, averaged over the range of  $ V_\mathrm{g}$ indicated by the white boxes in panels a, b, and c. \textbf{e} WL measured at $D/\epsilon_0=-0.4~$V/nm and  $V_\mathrm{g}=-97$~mV. The WL signal is clearly visible at the top of the BLG valence band in panels b and c even without averaging over different gate voltages (see orange and red arrows).
}
\label{f3}
\end{figure*}

The SOC tuning in BLG/TMD heterostructures relies on the tunable band gap of BLG under an applied out-of-plane  displacement field~\cite{McCann_2013, Slizovskiy2019Dec}.
As in gapped BLG at low charge carrier density, electrons and holes are hosted on distinct graphene layers~\cite{Young2011, Levitov2017, JFabianOct2017,Wang2021,Island2019, Masseroni2024Oct, Seiler2024Mar}, and since proximity-induced SOC acts only on the layer in direct contact with the TMD (i.e., the proximal layer L1 in Fig.~\ref{f1}a), either the conduction or the valence band is spin-split depending on the direction of the applied electric displacement field~\cite{Levitov2017, JFabianOct2017}, see example in Fig.~\ref{f1}b. 
A double-gated device geometry, incorporating both top and bottom gates (as schematically shown in Fig.~\ref{f1}a), enables independent control of the Fermi level $\mu_\text{F}$ and the band gap $E_\mathrm{g}$, thereby allowing charge carriers to be shifted between layers with strong and weak SOC. This tunability forms the fundamental operating principle of a spin–orbit valve, which has been studied both theoretically~\cite{JFabianOct2017, Levitov2017} and experimentally~\cite{Tiwari2021Mar}.

Proximity-induced SOC in BLG also affects the Berry phase accumulated when encircling one of the K-points in the Brillouin zone. In pristine BLG with negligible SOC, the Berry phase is  $\pm 2 \pi$, which leads to the observation of weak localization (WL) in diffusive, phase-coherent quantum transport~\cite{Gorbachev2007,Liao2010Oct,Engels2014Sep}. In contrast, when Rashba-type SOC is induced — for example, via proximity to a TMD such as WSe$_2$ — the Berry phase is reduced to  $\pm  \pi$~\cite{Zhai2022May}, resulting in weak anti-localization (WAL) ~\cite{Wang2015nov, Wang2016oct, Yang2016sep, Yang2017jul, Voelkl2017Sep, Zihlmann2018Feb, Afzal2018feb, Wakamura2018Mar, Jafarpisheh2018Dec, amann2021}.
The appearance of WAL is, in fact, commonly regarded as a hallmark of proximity-induced SOC in BLG-based heterostructures.
The situation becomes more complex when a band gap is opened, as the accumulated Berry phase then depends on the 
$k$-values that define the closed orbit. However, for orbits near the band edge, i.e. the K-points, the accumulated Berry phase approaches zero, with its detailed dependence determined by the specifics of the proximity-induced SOC and the spin-split band structure.
This gives rise to WL near the band edge, making it a sensitive probe for band structure properties.

In this work, we report on a high-quality BLG/\ch{WSe2} device that exhibit both high charge carrier mobility and excellent band gap tunability — indicative of minimal interfacial disorder. The device operates in a quasi-ballistic transport regime. We demonstrate that robust electrostatic control enables the formation of gate-defined lateral p–n–p cavities, which give rise to weak anti-localization features in the magneto-conductance. Furthermore, we observe pronounced weak localization that can be unambiguously attributed to the proximity-induced spin-split valence band of the graphene layer in contact with \ch{WSe2}. This provides direct evidence of the proximity-induced spin splitting probed by transport spectroscopy.

The device consists of a heterostructure of BLG/\ch{WSe2} encapsulated between two crystals of hexagonal boron nitride (hBN) equipped with a bottom gate made of graphite (Gr) and a narrow top gate made of gold (see schematic in Fig.~\ref{f1}a and Raman spectra of the layered stack in Supporting Information Section I). The graphite gate screens the disorder potential from the underlying SiO$_2$ substrate, allowing the opening of a clean band gap in BLG \cite{Icking2022Oct}. Flakes are selected by an automated flake search tool, where the thickness of the \ch{WSe2} flakes is determined via their optical contrast \cite{Uslu2023Jun,Uslu2024Dec}. Since the lateral size of the top gate is smaller than the bottom gate, only part of the BLG/\ch{WSe2} heterostructure is double gated, while the remaining parts -- which in the following will be called `leads' -- are only tuned by the bottom gate, see Figs.~\ref{f1}c and \ref{f1}d. This allows the realization of electrostatically-induced p-n junctions at the boundaries of the top-gated region in transport direction. For instance, the voltage configuration indicated in the schematics of Fig.~\ref{f1}c corresponds to a situation where the leads are p-doped while the double-gated region is n-doped and has a band gap $E_\mathrm{g}$ induced by the electric displacement field.

The low temperature, two-terminal resistance of such a device as function of the voltages applied to the top and bottom gate, $V_\mathrm{tg}$ and $V_\mathrm{bg}$, is shown in Fig.~\ref{f2}a. 
The diagonal feature of elevated resistance indicates the shift of the midgap energy of the BLG 
in the double-gated region as a function of $V_\mathrm{tg}$ and $V_\mathrm{bg}$, while the vertical feature
at $V_\mathrm{bg}= -0.12$\,V originates from the charge neutrality point of the BLG regions that are only tuned by the back gate. The broadening of the diagonal high-resistance feature with increasing $V_\mathrm{tg}$ and $V_\mathrm{bg}$ indicates the opening of a band gap in the double-gated BLG region~\cite{Icking2022Oct}. 
It is convenient to introduce the electric displacement field $D$ in the double gated region,   
$D=e\alpha_\mathrm{tg}\Big[\beta\big(V_\mathrm{bg}-V^0_\mathrm{bg}\big)-\big(V_\mathrm{tg}-V^0_\mathrm{tg}\big)\Big]/{2}$, 
and the effective gate voltage 
$V_\mathrm{g} = \beta\left( V_{bg} - V_\mathrm{bg}^0\right) + \left( V_\mathrm{tg} - V_\mathrm{tg}^0 \right)/(1 + \beta)$, where  
$\beta =\alpha_\mathrm{tg}/\alpha_\mathrm{bg}$ is the relative lever arm of the top and bottom gate, $\alpha_\mathrm{tg}$ the top-gate lever arm,  $V_\mathrm{tg, bg}^0$ account for the residual extrinsic doping of the BLG, and $e$ and $\varepsilon_0$ are the elementary charge and the vacuum permittivity, respectively~\cite{Icking2022Oct}. For the device shown in Fig.~\ref{f1}d, the relative lever-arm is $\beta=0.745$, which corresponds to the slope of the diagonal line of high resistance in Fig.~\ref{f2}a, while $\alpha_\mathrm{tg} = 5.9 \times 10^{15}$\,V$^{-1}$m$^{-2}$ is  extracted from quantum Hall measurements (see Supporting Information Section II). 
By plotting the resistance in terms of  $D$ and $V_\mathrm{g}$ (see Fig.~\ref{f2}b), it becomes apparent that the width of the high-resistance region increases linearly with $|D|$, as expected from the opening of a band gap in BLG (see black dashed lines in Fig.~\ref{f2}b)~\cite{Icking2022Oct}.

The resistance data furthermore present a distinct asymmetry with respect to $V_\mathrm{g}$ (see also line traces in Fig.~\ref{f2}d) which 
can be attributed to the formation of lateral p-n junctions at the position of the edges of the top gate, depending on the value of $V_\mathrm{g}$ and $D$. 
For instance, for the case of Fig.~\ref{f2}d ($D<0$), the region of low resistance for $V_\mathrm{g}<0$ corresponds to the situation in which the Fermi level, $\mu_\mathrm{F}$, is in the valence band of both the leads and of the double-gated central region; increasing $V_\mathrm{g}$, $\mu_F$ enters the band-gap of the double-gated region, which leads to a sharp increase of resistance of almost four orders of magnitude. 
For even larger values of $V_\mathrm{g}$, $\mu_F$ enters the conduction band of the double-gated region, resulting in the formation of a p-n-p cavity. 
The resistance drops but remains significantly higher than for negative voltages due to the enhanced dwell times of the charge carriers in the cavity itself. 
The same behaviour is observed for $D>0$ and $V_\mathrm{g}<0$, with the formation of a n-p-n junction.   

The asymmetry with respect to $V_\mathrm{g}$ is also observed in the finite-bias spectroscopy measurements presented in Fig.~\ref{f2}c.  In this type of measurement, the appearance of a diamond-shaped region of suppressed conductance  is the hallmark of the opening of a clean band gap in BLG~\cite{Icking2022Oct, Icking2024Sep}. The edges of the region of suppressed conductance correspond to the situation in which the electrochemical  potential in the leads is aligned with the band edges in the double-gated region~\cite{Icking2022Oct, Icking2024Sep}. For perfectly clean band gaps, the edges of the diamond region are expected to have a slope of two. 
For the device of Fig.~\ref{f1}d, the diamond shape is most pronounced at large displacement fields, where the slope of the outer edges reaches values of $\Delta V_\mathrm{sd}/\Delta V_\mathrm{g}=1.9$ for the valence band edge. This small deviation might be attributed to the presence of defects and trap states in the \ch{WSe2} layer. The presence of defect or subgap states in the \ch{WSe2} would also account for the observation  that the maximum resistance measured in this type of device is about two orders of magnitude smaller than that observed in BLG/hBN devices with graphite gates but without the \ch{WSe2} layer, where resistance values as high as 100~G$\Omega$ have been reported~\cite{Icking2022Oct}. Hopping transport through the trap states in \ch{WSe2} would, indeed, represent a parallel conduction channel and lower the overall maximum resistance of the BLG/\ch{WSe2} device compared to BLG/hBN devices.

The finite-bias spectroscopy measurements in Fig. ~\ref{f2}c serve as a direct probe of the magnitude of the electrostatically induced band gap in BLG, $E_\mathrm{g}$, which is directly proportional to the maximal extension of the diamond along the $V_\mathrm{sd}$ axis. In Fig.~\ref{f2}e, we show respective value of $E_\mathrm{g}$ for both positive (full symbols) and negative (empty symbols) displacement fields. While the measured values of $E_\mathrm{g}$ are offset by about $\sim12$~meV from what is theoretically predicted for BLG (dashed gray line), the data indicate a good band gap tunability in the BLG/\ch{WSe2} heterostructure. 
The offset can be attributed to residual  disorder~\cite{Icking2022Oct}, but may also originate partially from SOC-induced spin-split bands~\cite{Gmitra2017Oct}. 

The good tunability of the band gap, $E_\mathrm{g}$, is a strong indication of
low disorder in the heterostructure~\cite{Icking2022Oct}, which is also reflected by the high charge carrier mobility. For the device of Fig.~\ref{f1}d we extract $\mu_{e} \approx 260,000$\,cm$^{2}$V$^{-1}$s$^{-1}$ for electron transport  at  $T=60$~mK (see Supporting Information Section III). This results in a mean free path $l_\mathrm{m} \approx 1.0 - 2.0 \,\upmu$m, which is comparable with the length (i.e., the lateral extent in the transport direction) of the top gate, $l=2\,\upmu$m, implying the device is in a quasi-ballistic transport regime. 

To probe the proximity-induced SOC in BLG, we next discuss magneto-conductance measurements as a function of $V_\mathrm{g}$ for different values of the applied displacement field. In diffusive samples, it has been shown that the proximity-induced SOC results in the appearance of a weak antilocalization peak at low magnetic field~\cite{Yang2016sep, Wang2016oct, Afzal2018feb}.  Since our sample is quasi-ballistic, we expect to observe WAL only in gate configurations that correspond to the formation of a p-n-p cavity (i.e. for $D<0$ and $V_g>0$), or of a n-p-n cavity (i.e. for $D>0$ and $V_g<0$), where quantum interference effects can arise due to internal reflections at the p-n junctions which results in extended dwell times.  

\begin{figure*}[hbt!]
	\centering
\includegraphics[draft=false,keepaspectratio=true,clip,width=0.96\linewidth]{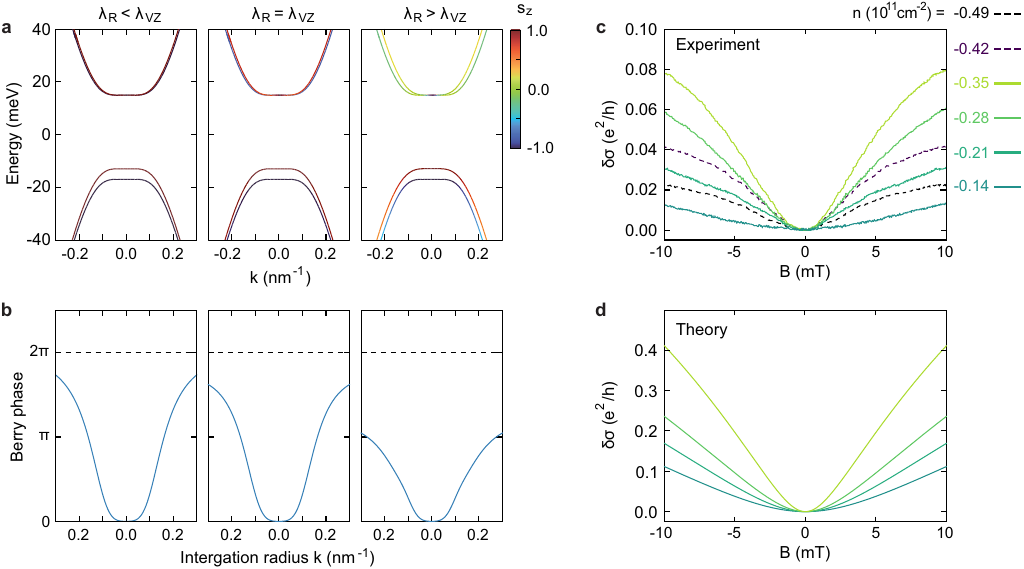}
\caption[Fig04]{\textbf{a} Calculated low-energy bands of a BLG/\ch{WSe2} heterostructure for different ratios between the Rashba SOC strength $\lambda_\text{R}$ and the Valley-Zeeman SOC strength $\lambda_\text{VZ}$ at the K point ($\tau=1$). The color scale denotes the out-of-plane spin orientation.
\textbf{b} The corresponding Berry phase of the highest spin-split valence band, calculated around a closed contour in reciprocal space at a fixed momentum magnitude $k$ with the center being the K point. Close to the band edge, the Berry phase is zero for the upper valence band, suggesting that WAL is completely suppressed when only one band is occupied~\cite{Kechedzhi2007Sep, Wang2016oct, Yang2016sep, Lu2011}. Further into the band, the Berry phase approaches $2\pi$ if $\lambda_\text{R}<\lambda_\text{VZ}$, or if $\lambda_\text{R}=\lambda_\text{VZ}$, and approaches $\pi$ if $\lambda_\text{R}>\lambda_\text{VZ}$. This behavior is seen at the K' point except for a sign switch in the Berry phase. \textbf{c} Normalized magneto-conductance $\delta\sigma$ extracted from a
$\Delta V_\text{g}$ regime close to the valence band edge (see orange outline in Fig.\ref{f3}b). \textbf{d }Calculated WL signal within the spin-split bands, assuming that scattering and spin relaxation only occur at the edges of the top-gate-defined cavity and the BLG. The results show that for increasing hole density $|n|$ the width of the WL decreases, which is consistent with our measurements for $n\in[-0.14,-0.35]\times10^{11}$\,cm$^{-2}$. For higher $|n|$, we get contributions from the other spin-split band, resulting in a suppression of the WL signal.}
\label{f4}
\end{figure*}

Magneto-transport measurements at different values of the applied displacement field are presented in Fig.~\ref{f3} for $D<0$. Around $V_\mathrm{g}\approx 0$~V, the conductance is strongly suppressed and does not depend on $B$, indicating that the Fermi level ($\mu_\mathrm{F}$) is in the band gap of the double-gated region. Outside this gap region, we observe a complicated pattern of universal conductance fluctuations (see Supporting Information Section IV) as well as the emergence of a WAL peak at positive $V_\mathrm{g}$, when $\mu_\mathrm{F}$ is in the conduction band of the double-gated region forming a p-n-p cavity. This feature is best visualized by averaging different conductance traces measured over a range of gate voltages $V_\mathrm{g}$, see  Fig.~\ref{f3}d. The respective gate voltage ranges are indicated by white boxes for the different displacement fields in Figs.~\ref{f3}a to~\ref{f3}c.  
Interestingly, WAL is well developed only for large  displacement fields $|D|$ (see Supporting Information Section V), which might be a good indication that the p-n-p cavity plays an important role. From the width of the WAL peaks, we can estimate the phase coherence length~\cite{Kechedzhi2007Sep, ilic2019may}, obtaining  $l_\phi\approx\sqrt{h/(2e\Delta B_\mathrm{FWHM})}\approx 3$~$\upmu$m, which fulfills $l_\phi \gtrapprox l_\mathrm{m}$, the condition for observing quantum interference effects~\cite{BluhmBuch2019}. 

The most remarkable feature of the data of Fig.~\ref{f3} is, however, the existence of a narrow and sharply defined range of $V_\mathrm{g}$ at the top of the valence band, where the conductance exhibits a pronounced weak localization dip (see orange and red arrows in Figs.~\ref{f3}b,c). 
The amplitude and the width of the WL signal exceeds those of the WAL signals by up to an order of magnitude (Fig.~\ref{f3}e).  
We attribute this WL signal to transport through a single spin-polarized band in the double-gated region, which in turn results from the splitting of the valence band in BLG due to proximity-induced SOC. It has been shown, both in experiment~\cite{Island2019} and theory~\cite{Levitov2017, Gmitra2017Oct}, that proximity-induced SOC leads to a spin-split valence band for $D<0$, as sketched in Fig.~\ref{f1}b. 
In our device, this spin splitting leads to a conductance mismatch between the central double-gated region and the adjacent BLG regions in the range of gate voltages when only the upper valence band in the double-gated region is relevant for transport.  The range of $V_g$ in which WL can be observed is therefore directly related to band splitting induced by SOC, $\Delta_\mathrm{SOC}$. 
By converting the  $V_\mathrm{g}$ range where we observe WL into a corresponding energy shift, we roughly estimate $\Delta_\mathrm{SOC} \approx2$\,meV for $D=-300~$mV/nm and $\Delta_\mathrm{SOC} \approx3$\,meV for $D=-400~$mV/nm (Supporting Information Section VI). These values are consistent with theoretical predictions~\cite{Gmitra2017Oct, Levitov2017, Zollner2020Nov, Zollner2021Aug} and experimental data of proximity-induced band splitting~\cite{Island2019, Wang2021}.

To interpret our experimental findings, it is crucial to consider the nature of carrier transport near the band edge of proximity-coupled gapped BLG. In this regime, the mean free path is significantly reduced due to two main factors: (i) enhanced scattering from impurity states and disorder introduced by the adjacent WSe$_2$ layer, and (ii) the relatively low Fermi velocity $v_\mathrm{F}$ (or, in other terms, the high effective mass) near the band edge of gapped BLG. Both effects drive the system into a more diffusive transport regime.
In addition, we note that the band mismatch between the double-gated BLG region (with one valence band) and the lead regions (with two valence bands) may also create an effective cavity, which increases the dwell time and thereby enhances the WL signal.

Importantly, the observed WL signatures near the valence band edge — where only one valence band is relevant — are consistent with expectations based on Berry phase considerations.
In Figs.~\ref{f4}a,b we show calculations of the band structure of the proximity-coupled WSe$_2$/BLG system for different SOC strength and the corresponding accumulated Berry phase for circles around the K-point (for details see Supporting Information Sections VII and VIII).
To adjust the band gap and valence band spin splitting, we choose a layer potential difference of $15$ meV and the valley-Zeeman SOC strength $\lvz = 2$~meV. The different panels of Figs.~\ref{f4}a,b correspond to different strengths of Rashba SOC, namely $\lr = \lvz / 5$, $\lr = \lvz$, and $\lr = 5\lvz$ from left to right. In Fig.~\ref{f4}a, we see that independent of the Rashba SOC strength, the valence band is always strongly spin-polarized along the $z$-axis (out-of-plane direction) near the band edge.
In Fig.~\ref{f4}b, we plot (corresponding to the panels in Fig.~\ref{f4}a) the accumulated Berry phase as a function of the integration radius $k$ 
(for more information on the calculations see Supporting Information Section VIII). Crucially, the resulting Berry phase is always close to $0$ near the band edge (i.e., K-point), giving rise to a pronounced WL instead of a WAL signal when only the upper valence band is occupied~\cite{Lu2011}.
This finding is independent of the relative Rashba and valley-Zeeman strength and is consistent with our measurements. The same argumentation holds for the K'-point without a loss of generality except for a sign change of the Berry phase for increasing~ $k$. 

Having established that the WL signal is consistent with band structure calculations, we now examine how the WL signal evolves with the hole density. As shown in Fig.~\ref{f4}c, the amplitude of the WL signal rises sharply at low hole densities near the top of the valence band, peaks at $n = -0.35\times10^{11}$\,cm$^{-2}$, and then decreases again at higher hole densities.
We model this behavior by considering a general expression for WL/WAL in BLG~\cite{amann2021} (for more details on the model, see Supporting Information Section IX). 
The scattering parameters are estimated based on the cavity dimensions, the diffusion constant, and the Fermi velocity 
$v_\text{F}(n)$, which we extract from band structure calculations.
Figure~\ref{f4}d shows calculated WL curves for 
carrier densities in the range of $-0.14$ to $-0.35\times10^{11}$\,cm$^{-2}$  based on the scattering model. The increasing WL amplitude with hole density matches well the experimental trend in Fig.~\ref{f4}c.
Since our model includes only the upper valence band, the WL amplitude increases with hole density across the entire range. In contrast, the measured WL amplitude drops sharply when the second valence band is populated (see dashed curves in Fig.~\ref{f4}c). Note that at low hole densities, both experiment and simulation show a clear broadening of the WL curves. Our model attributes this to a reduced diffusion coefficient, which scales with 
$v_\text{F}^2/ 2$and thus decreases with lowering hole density.

In summary, we have presented a double-gated BLG/\ch{WSe2}  heterostructure that combines high mobility, excellent electrostatic control, and clear signatures of proximity-induced spin–orbit coupling in phase-coherent transport. The use of graphite gates enables clean and tunable band gaps, almost comparable to those in pristine BLG, and allows full access to both conduction and valence bands. Notably, no screening effects from the \ch{WSe2}  layer are observed, indicating a well-behaved interface.

Magneto-transport measurements reveal a transition from weak anti-localization to weak localization as the Fermi level is tuned into the upper valence band. This behavior is consistent with theoretical expectations: near the band edge, the Berry phase drops towards zero, independent of the relative strength of Rashba and valley-Zeeman spin–orbit coupling. The observed trends in the WL amplitude as a function of hole density are captured by a scattering model, using input parameters derived from band structure calculations.
Taken together, these findings provide spectroscopic evidence for proximity-induced spin split bands in BLG, probed by phase-coherent transport. The combination of low disorder, sharp band edges, and precise gate control establishes a robust platform for spin-sensitive mesoscopic devices. Therefore such systems open the door to exploring gate-defined quantum point contacts, spin filters, and  quantum dots in bilayer graphene with tunable spin-orbit coupling, as also demonstrated very recently~\cite{Gerber2025Apr,Dulisch2025Apr}.
\newline

\textbf{Acknowledgements} The authors thank S.~Trellenkamp, F.~Lentz and D.~Neumaier for their support in device fabrication. We also thank
J.~Eroms, J.~Amann, and J.~Fabian for discussions about the nature of WAL, K.~Zollner for discussions about proximity effects in BLG on \ch{WSe2}, and F.~Haupt for help on the manuscript.
This project has received funding from the European Union's Horizon 2020 research and innovation programme under grant agreement No. 881603 (Graphene Flagship) and from the European Research Council (ERC) under grant agreement No. 820254, the Deutsche Forschungsgemeinschaft (DFG, German Research Foundation) under Germany's Excellence Strategy - Cluster of Excellence Matter and Light for Quantum Computing (ML4Q) EXC 2004/1 - 390534769 and by the Helmholtz Nano Facility~\cite{Albrecht2017May}. K.W. and T.T. acknowledge support from JSPS KAKENHI (Grant Numbers 19H05790, 20H00354 and 21H05233).
ICN2 is funded by the CERCA programme / Generalitat de Catalunya, and is supported by the Severo Ochoa Centres of Excellence programme, Grant CEX2021-001214-S, funded by MCIN/AEI/10.13039.501100011033.

\clearpage


\begin{thebibliography}{50}%
\makeatletter
\providecommand \@ifxundefined [1]{%
 \@ifx{#1\undefined}
}%
\providecommand \@ifnum [1]{%
 \ifnum #1\expandafter \@firstoftwo
 \else \expandafter \@secondoftwo
 \fi
}%
\providecommand \@ifx [1]{%
 \ifx #1\expandafter \@firstoftwo
 \else \expandafter \@secondoftwo
 \fi
}%
\providecommand \natexlab [1]{#1}%
\providecommand \enquote  [1]{``#1''}%
\providecommand \bibnamefont  [1]{#1}%
\providecommand \bibfnamefont [1]{#1}%
\providecommand \citenamefont [1]{#1}%
\providecommand \href@noop [0]{\@secondoftwo}%
\providecommand \href [0]{\begingroup \@sanitize@url \@href}%
\providecommand \@href[1]{\@@startlink{#1}\@@href}%
\providecommand \@@href[1]{\endgroup#1\@@endlink}%
\providecommand \@sanitize@url [0]{\catcode `\\12\catcode `\$12\catcode `\&12\catcode `\#12\catcode `\^12\catcode `\_12\catcode `\%12\relax}%
\providecommand \@@startlink[1]{}%
\providecommand \@@endlink[0]{}%
\providecommand \url  [0]{\begingroup\@sanitize@url \@url }%
\providecommand \@url [1]{\endgroup\@href {#1}{\urlprefix }}%
\providecommand \urlprefix  [0]{URL }%
\providecommand \Eprint [0]{\href }%
\providecommand \doibase [0]{http://dx.doi.org/}%
\providecommand \selectlanguage [0]{\@gobble}%
\providecommand \bibinfo  [0]{\@secondoftwo}%
\providecommand \bibfield  [0]{\@secondoftwo}%
\providecommand \translation [1]{[#1]}%
\providecommand \BibitemOpen [0]{}%
\providecommand \bibitemStop [0]{}%
\providecommand \bibitemNoStop [0]{.\EOS\space}%
\providecommand \EOS [0]{\spacefactor3000\relax}%
\providecommand \BibitemShut  [1]{\csname bibitem#1\endcsname}%
\let\auto@bib@innerbib\@empty
%</preamble>
\bibitem [{\citenamefont {Gmitra}\ and\ \citenamefont {Fabian}(2015)}]{Gmitra2015oct}%
  \BibitemOpen
  \bibfield  {author} {\bibinfo {author} {\bibfnamefont {M.}~\bibnamefont {Gmitra}}\ and\ \bibinfo {author} {\bibfnamefont {J.}~\bibnamefont {Fabian}},\ }\href {\doibase 10.1103/PhysRevB.92.155403} {\bibfield  {journal} {\bibinfo  {journal} {Phys. Rev. B}\ }\textbf {\bibinfo {volume} {92}},\ \bibinfo {pages} {155403} (\bibinfo {year} {2015})}\BibitemShut {NoStop}%
\bibitem [{\citenamefont {Gmitra}\ \emph {et~al.}(2016)\citenamefont {Gmitra}, \citenamefont {Kochan}, \citenamefont {H\"ogl},\ and\ \citenamefont {Fabian}}]{Gmitra2016apr}%
  \BibitemOpen
  \bibfield  {author} {\bibinfo {author} {\bibfnamefont {M.}~\bibnamefont {Gmitra}}, \bibinfo {author} {\bibfnamefont {D.}~\bibnamefont {Kochan}}, \bibinfo {author} {\bibfnamefont {P.}~\bibnamefont {H\"ogl}}, \ and\ \bibinfo {author} {\bibfnamefont {J.}~\bibnamefont {Fabian}},\ }\href {\doibase 10.1103/PhysRevB.93.155104} {\bibfield  {journal} {\bibinfo  {journal} {Phys. Rev. B}\ }\textbf {\bibinfo {volume} {93}},\ \bibinfo {pages} {155104} (\bibinfo {year} {2016})}\BibitemShut {NoStop}%
\bibitem [{\citenamefont {Wang}\ \emph {et~al.}(2015)\citenamefont {Wang}, \citenamefont {Ki}, \citenamefont {Chen}, \citenamefont {Berger}, \citenamefont {MacDonald},\ and\ \citenamefont {Morpurgo}}]{Wang2015nov}%
  \BibitemOpen
  \bibfield  {author} {\bibinfo {author} {\bibfnamefont {Z.}~\bibnamefont {Wang}}, \bibinfo {author} {\bibfnamefont {D.-K.}\ \bibnamefont {Ki}}, \bibinfo {author} {\bibfnamefont {H.}~\bibnamefont {Chen}}, \bibinfo {author} {\bibfnamefont {H.}~\bibnamefont {Berger}}, \bibinfo {author} {\bibfnamefont {A.~H.}\ \bibnamefont {MacDonald}}, \ and\ \bibinfo {author} {\bibfnamefont {A.~F.}\ \bibnamefont {Morpurgo}},\ }\href {\doibase 10.1038/ncomms9339} {\bibfield  {journal} {\bibinfo  {journal} {Nature Communications}\ }\textbf {\bibinfo {volume} {6}},\ \bibinfo {pages} {8339} (\bibinfo {year} {2015})}\BibitemShut {NoStop}%
\bibitem [{\citenamefont {Wang}\ \emph {et~al.}(2016)\citenamefont {Wang}, \citenamefont {Ki}, \citenamefont {Khoo}, \citenamefont {Mauro}, \citenamefont {Berger}, \citenamefont {Levitov},\ and\ \citenamefont {Morpurgo}}]{Wang2016oct}%
  \BibitemOpen
  \bibfield  {author} {\bibinfo {author} {\bibfnamefont {Z.}~\bibnamefont {Wang}}, \bibinfo {author} {\bibfnamefont {D.-K.}\ \bibnamefont {Ki}}, \bibinfo {author} {\bibfnamefont {J.~Y.}\ \bibnamefont {Khoo}}, \bibinfo {author} {\bibfnamefont {D.}~\bibnamefont {Mauro}}, \bibinfo {author} {\bibfnamefont {H.}~\bibnamefont {Berger}}, \bibinfo {author} {\bibfnamefont {L.~S.}\ \bibnamefont {Levitov}}, \ and\ \bibinfo {author} {\bibfnamefont {A.~F.}\ \bibnamefont {Morpurgo}},\ }\href {\doibase 10.1103/PhysRevX.6.041020} {\bibfield  {journal} {\bibinfo  {journal} {Phys. Rev. X}\ }\textbf {\bibinfo {volume} {6}},\ \bibinfo {pages} {041020} (\bibinfo {year} {2016})}\BibitemShut {NoStop}%
\bibitem [{\citenamefont {Yang}\ \emph {et~al.}(2017)\citenamefont {Yang}, \citenamefont {Lohmann}, \citenamefont {Barroso}, \citenamefont {Liao}, \citenamefont {Lin}, \citenamefont {Liu}, \citenamefont {Bartels}, \citenamefont {Watanabe}, \citenamefont {Taniguchi},\ and\ \citenamefont {Shi}}]{Yang2017jul}%
  \BibitemOpen
  \bibfield  {author} {\bibinfo {author} {\bibfnamefont {B.}~\bibnamefont {Yang}}, \bibinfo {author} {\bibfnamefont {M.}~\bibnamefont {Lohmann}}, \bibinfo {author} {\bibfnamefont {D.}~\bibnamefont {Barroso}}, \bibinfo {author} {\bibfnamefont {I.}~\bibnamefont {Liao}}, \bibinfo {author} {\bibfnamefont {Z.}~\bibnamefont {Lin}}, \bibinfo {author} {\bibfnamefont {Y.}~\bibnamefont {Liu}}, \bibinfo {author} {\bibfnamefont {L.}~\bibnamefont {Bartels}}, \bibinfo {author} {\bibfnamefont {K.}~\bibnamefont {Watanabe}}, \bibinfo {author} {\bibfnamefont {T.}~\bibnamefont {Taniguchi}}, \ and\ \bibinfo {author} {\bibfnamefont {J.}~\bibnamefont {Shi}},\ }\href {\doibase 10.1103/PhysRevB.96.041409} {\bibfield  {journal} {\bibinfo  {journal} {Phys. Rev. B}\ }\textbf {\bibinfo {volume} {96}},\ \bibinfo {pages} {041409} (\bibinfo {year} {2017})}\BibitemShut {NoStop}%
\bibitem [{\citenamefont {V\"olkl}\ \emph {et~al.}(2017)\citenamefont {V\"olkl}, \citenamefont {Rockinger}, \citenamefont {Drienovsky}, \citenamefont {Watanabe}, \citenamefont {Taniguchi}, \citenamefont {Weiss},\ and\ \citenamefont {Eroms}}]{Voelkl2017Sep}%
  \BibitemOpen
  \bibfield  {author} {\bibinfo {author} {\bibfnamefont {T.}~\bibnamefont {V\"olkl}}, \bibinfo {author} {\bibfnamefont {T.}~\bibnamefont {Rockinger}}, \bibinfo {author} {\bibfnamefont {M.}~\bibnamefont {Drienovsky}}, \bibinfo {author} {\bibfnamefont {K.}~\bibnamefont {Watanabe}}, \bibinfo {author} {\bibfnamefont {T.}~\bibnamefont {Taniguchi}}, \bibinfo {author} {\bibfnamefont {D.}~\bibnamefont {Weiss}}, \ and\ \bibinfo {author} {\bibfnamefont {J.}~\bibnamefont {Eroms}},\ }\href {\doibase 10.1103/PhysRevB.96.125405} {\bibfield  {journal} {\bibinfo  {journal} {Phys. Rev. B}\ }\textbf {\bibinfo {volume} {96}},\ \bibinfo {pages} {125405} (\bibinfo {year} {2017})}\BibitemShut {NoStop}%
\bibitem [{\citenamefont {Zihlmann}\ \emph {et~al.}(2018)\citenamefont {Zihlmann}, \citenamefont {Cummings}, \citenamefont {Garcia}, \citenamefont {Kedves}, \citenamefont {Watanabe}, \citenamefont {Taniguchi}, \citenamefont {Sch\"onenberger},\ and\ \citenamefont {Makk}}]{Zihlmann2018Feb}%
  \BibitemOpen
  \bibfield  {author} {\bibinfo {author} {\bibfnamefont {S.}~\bibnamefont {Zihlmann}}, \bibinfo {author} {\bibfnamefont {A.~W.}\ \bibnamefont {Cummings}}, \bibinfo {author} {\bibfnamefont {J.~H.}\ \bibnamefont {Garcia}}, \bibinfo {author} {\bibfnamefont {M.}~\bibnamefont {Kedves}}, \bibinfo {author} {\bibfnamefont {K.}~\bibnamefont {Watanabe}}, \bibinfo {author} {\bibfnamefont {T.}~\bibnamefont {Taniguchi}}, \bibinfo {author} {\bibfnamefont {C.}~\bibnamefont {Sch\"onenberger}}, \ and\ \bibinfo {author} {\bibfnamefont {P.}~\bibnamefont {Makk}},\ }\href {\doibase 10.1103/PhysRevB.97.075434} {\bibfield  {journal} {\bibinfo  {journal} {Phys. Rev. B}\ }\textbf {\bibinfo {volume} {97}},\ \bibinfo {pages} {075434} (\bibinfo {year} {2018})}\BibitemShut {NoStop}%
\bibitem [{\citenamefont {Wakamura}\ \emph {et~al.}(2018)\citenamefont {Wakamura}, \citenamefont {Reale}, \citenamefont {Palczynski}, \citenamefont {Gu\'eron}, \citenamefont {Mattevi},\ and\ \citenamefont {Bouchiat}}]{Wakamura2018Mar}%
  \BibitemOpen
  \bibfield  {author} {\bibinfo {author} {\bibfnamefont {T.}~\bibnamefont {Wakamura}}, \bibinfo {author} {\bibfnamefont {F.}~\bibnamefont {Reale}}, \bibinfo {author} {\bibfnamefont {P.}~\bibnamefont {Palczynski}}, \bibinfo {author} {\bibfnamefont {S.}~\bibnamefont {Gu\'eron}}, \bibinfo {author} {\bibfnamefont {C.}~\bibnamefont {Mattevi}}, \ and\ \bibinfo {author} {\bibfnamefont {H.}~\bibnamefont {Bouchiat}},\ }\href {\doibase 10.1103/PhysRevLett.120.106802} {\bibfield  {journal} {\bibinfo  {journal} {Phys. Rev. Lett.}\ }\textbf {\bibinfo {volume} {120}},\ \bibinfo {pages} {106802} (\bibinfo {year} {2018})}\BibitemShut {NoStop}%
\bibitem [{\citenamefont {Banszerus}\ \emph {et~al.}(2019)\citenamefont {Banszerus}, \citenamefont {Sohier}, \citenamefont {Epping}, \citenamefont {Winkler}, \citenamefont {Libisch}, \citenamefont {Haupt}, \citenamefont {Watanabe}, \citenamefont {Taniguchi}, \citenamefont {M{\ifmmode\ddot{u}\else\"{u}\fi}ller-Caspary}, \citenamefont {Marzari}, \citenamefont {Mauri}, \citenamefont {Beschoten},\ and\ \citenamefont {Stampfer}}]{Banszerus2019Sep}%
  \BibitemOpen
  \bibfield  {author} {\bibinfo {author} {\bibfnamefont {L.}~\bibnamefont {Banszerus}}, \bibinfo {author} {\bibfnamefont {T.}~\bibnamefont {Sohier}}, \bibinfo {author} {\bibfnamefont {A.}~\bibnamefont {Epping}}, \bibinfo {author} {\bibfnamefont {F.}~\bibnamefont {Winkler}}, \bibinfo {author} {\bibfnamefont {F.}~\bibnamefont {Libisch}}, \bibinfo {author} {\bibfnamefont {F.}~\bibnamefont {Haupt}}, \bibinfo {author} {\bibfnamefont {K.}~\bibnamefont {Watanabe}}, \bibinfo {author} {\bibfnamefont {T.}~\bibnamefont {Taniguchi}}, \bibinfo {author} {\bibfnamefont {K.}~\bibnamefont {M{\ifmmode\ddot{u}\else\"{u}\fi}ller-Caspary}}, \bibinfo {author} {\bibfnamefont {N.}~\bibnamefont {Marzari}}, \bibinfo {author} {\bibfnamefont {F.}~\bibnamefont {Mauri}}, \bibinfo {author} {\bibfnamefont {B.}~\bibnamefont {Beschoten}}, \ and\ \bibinfo {author} {\bibfnamefont {C.}~\bibnamefont {Stampfer}},\ }\href {\doibase 10.48550/arXiv.1909.09523} {\bibfield  {journal} {\bibinfo  {journal} {arXiv}\ } (\bibinfo {year} {2019}),\
  10.48550/arXiv.1909.09523},\ \Eprint {http://arxiv.org/abs/1909.09523} {1909.09523} \BibitemShut {NoStop}%
\bibitem [{\citenamefont {F{\ifmmode\ddot{u}\else\"{u}\fi}l{\ifmmode\ddot{o}\else\"{o}\fi}p}\ \emph {et~al.}(2021)\citenamefont {F{\ifmmode\ddot{u}\else\"{u}\fi}l{\ifmmode\ddot{o}\else\"{o}\fi}p}, \citenamefont {M{\ifmmode\acute{a}\else\'{a}\fi}rffy}, \citenamefont {Zihlmann}, \citenamefont {Gmitra}, \citenamefont {T{\ifmmode\acute{o}\else\'{o}\fi}v{\ifmmode\acute{a}\else\'{a}\fi}ri}, \citenamefont {Szentp{\ifmmode\acute{e}\else\'{e}\fi}teri}, \citenamefont {Kedves}, \citenamefont {Watanabe}, \citenamefont {Taniguchi}, \citenamefont {Fabian}, \citenamefont {Sch{\ifmmode\ddot{o}\else\"{o}\fi}nenberger}, \citenamefont {Makk},\ and\ \citenamefont {Csonka}}]{Fulop2021Sep}%
  \BibitemOpen
  \bibfield  {author} {\bibinfo {author} {\bibfnamefont {B.}~\bibnamefont {F{\ifmmode\ddot{u}\else\"{u}\fi}l{\ifmmode\ddot{o}\else\"{o}\fi}p}}, \bibinfo {author} {\bibfnamefont {A.}~\bibnamefont {M{\ifmmode\acute{a}\else\'{a}\fi}rffy}}, \bibinfo {author} {\bibfnamefont {S.}~\bibnamefont {Zihlmann}}, \bibinfo {author} {\bibfnamefont {M.}~\bibnamefont {Gmitra}}, \bibinfo {author} {\bibfnamefont {E.}~\bibnamefont {T{\ifmmode\acute{o}\else\'{o}\fi}v{\ifmmode\acute{a}\else\'{a}\fi}ri}}, \bibinfo {author} {\bibfnamefont {B.}~\bibnamefont {Szentp{\ifmmode\acute{e}\else\'{e}\fi}teri}}, \bibinfo {author} {\bibfnamefont {M.}~\bibnamefont {Kedves}}, \bibinfo {author} {\bibfnamefont {K.}~\bibnamefont {Watanabe}}, \bibinfo {author} {\bibfnamefont {T.}~\bibnamefont {Taniguchi}}, \bibinfo {author} {\bibfnamefont {J.}~\bibnamefont {Fabian}}, \bibinfo {author} {\bibfnamefont {C.}~\bibnamefont {Sch{\ifmmode\ddot{o}\else\"{o}\fi}nenberger}}, \bibinfo {author} {\bibfnamefont {P.}~\bibnamefont {Makk}}, \ and\ \bibinfo {author}
  {\bibfnamefont {S.}~\bibnamefont {Csonka}},\ }\href {\doibase 10.1038/s41699-021-00262-9} {\bibfield  {journal} {\bibinfo  {journal} {npj 2D Mater. Appl.}\ }\textbf {\bibinfo {volume} {5}},\ \bibinfo {pages} {1} (\bibinfo {year} {2021})}\BibitemShut {NoStop}%
\bibitem [{\citenamefont {Tiwari}\ \emph {et~al.}(2022)\citenamefont {Tiwari}, \citenamefont {Jat}, \citenamefont {Udupa}, \citenamefont {Narang}, \citenamefont {Watanabe}, \citenamefont {Taniguchi}, \citenamefont {Sen},\ and\ \citenamefont {Bid}}]{Tiwari2022Oct}%
  \BibitemOpen
  \bibfield  {author} {\bibinfo {author} {\bibfnamefont {P.}~\bibnamefont {Tiwari}}, \bibinfo {author} {\bibfnamefont {M.~K.}\ \bibnamefont {Jat}}, \bibinfo {author} {\bibfnamefont {A.}~\bibnamefont {Udupa}}, \bibinfo {author} {\bibfnamefont {D.~S.}\ \bibnamefont {Narang}}, \bibinfo {author} {\bibfnamefont {K.}~\bibnamefont {Watanabe}}, \bibinfo {author} {\bibfnamefont {T.}~\bibnamefont {Taniguchi}}, \bibinfo {author} {\bibfnamefont {D.}~\bibnamefont {Sen}}, \ and\ \bibinfo {author} {\bibfnamefont {A.}~\bibnamefont {Bid}},\ }\href {\doibase 10.1038/s41699-022-00348-y} {\bibfield  {journal} {\bibinfo  {journal} {npj 2D Mater. Appl.}\ }\textbf {\bibinfo {volume} {6}},\ \bibinfo {pages} {1} (\bibinfo {year} {2022})}\BibitemShut {NoStop}%
\bibitem [{\citenamefont {Bisswanger}\ \emph {et~al.}(2025)\citenamefont {Bisswanger}, \citenamefont {Schmidt}, \citenamefont {Volmer}, \citenamefont {Stampfer},\ and\ \citenamefont {Beschoten}}]{Bisswanger2025May}%
  \BibitemOpen
  \bibfield  {author} {\bibinfo {author} {\bibfnamefont {T.}~\bibnamefont {Bisswanger}}, \bibinfo {author} {\bibfnamefont {A.}~\bibnamefont {Schmidt}}, \bibinfo {author} {\bibfnamefont {F.}~\bibnamefont {Volmer}}, \bibinfo {author} {\bibfnamefont {C.}~\bibnamefont {Stampfer}}, \ and\ \bibinfo {author} {\bibfnamefont {B.}~\bibnamefont {Beschoten}},\ }\href {\doibase 10.48550/arXiv.2505.18990} {\bibfield  {journal} {\bibinfo  {journal} {arXiv}\ } (\bibinfo {year} {2025}),\ 10.48550/arXiv.2505.18990},\ \Eprint {http://arxiv.org/abs/2505.18990} {2505.18990} \BibitemShut {NoStop}%
\bibitem [{\citenamefont {Garcia}\ \emph {et~al.}(2018)\citenamefont {Garcia}, \citenamefont {Vila}, \citenamefont {Cummings},\ and\ \citenamefont {Roche}}]{Garcia2018}%
  \BibitemOpen
  \bibfield  {author} {\bibinfo {author} {\bibfnamefont {J.~H.}\ \bibnamefont {Garcia}}, \bibinfo {author} {\bibfnamefont {M.}~\bibnamefont {Vila}}, \bibinfo {author} {\bibfnamefont {A.~W.}\ \bibnamefont {Cummings}}, \ and\ \bibinfo {author} {\bibfnamefont {S.}~\bibnamefont {Roche}},\ }\href {\doibase 10.1039/C7CS00864C} {\bibfield  {journal} {\bibinfo  {journal} {Chem. Soc. Rev.}\ }\textbf {\bibinfo {volume} {47}},\ \bibinfo {pages} {3359} (\bibinfo {year} {2018})}\BibitemShut {NoStop}%
\bibitem [{\citenamefont {Ahn}(2007)}]{Ahn2020}%
  \BibitemOpen
  \bibfield  {author} {\bibinfo {author} {\bibfnamefont {E.~C.}\ \bibnamefont {Ahn}},\ }\href {\doibase 10.1038/s41699-020-0152-0} {\bibfield  {journal} {\bibinfo  {journal} {npj 2D Materials and Applications}\ }\textbf {\bibinfo {volume} {4}},\ \bibinfo {pages} {17} (\bibinfo {year} {2007})}\BibitemShut {NoStop}%
\bibitem [{\citenamefont {Cummings}\ \emph {et~al.}(2017)\citenamefont {Cummings}, \citenamefont {Garcia}, \citenamefont {Fabian},\ and\ \citenamefont {Roche}}]{Cummings2017Nov}%
  \BibitemOpen
  \bibfield  {author} {\bibinfo {author} {\bibfnamefont {A.~W.}\ \bibnamefont {Cummings}}, \bibinfo {author} {\bibfnamefont {J.~H.}\ \bibnamefont {Garcia}}, \bibinfo {author} {\bibfnamefont {J.}~\bibnamefont {Fabian}}, \ and\ \bibinfo {author} {\bibfnamefont {S.}~\bibnamefont {Roche}},\ }\href {\doibase 10.1103/PhysRevLett.119.206601} {\bibfield  {journal} {\bibinfo  {journal} {Phys. Rev. Lett.}\ }\textbf {\bibinfo {volume} {119}},\ \bibinfo {pages} {206601} (\bibinfo {year} {2017})}\BibitemShut {NoStop}%
\bibitem [{\citenamefont {Ingla-Ayn\'es}\ \emph {et~al.}(2021)\citenamefont {Ingla-Ayn\'es}, \citenamefont {Herling}, \citenamefont {Fabian}, \citenamefont {Hueso},\ and\ \citenamefont {Casanova}}]{Ingla2021Jul}%
  \BibitemOpen
  \bibfield  {author} {\bibinfo {author} {\bibfnamefont {J.}~\bibnamefont {Ingla-Ayn\'es}}, \bibinfo {author} {\bibfnamefont {F.}~\bibnamefont {Herling}}, \bibinfo {author} {\bibfnamefont {J.}~\bibnamefont {Fabian}}, \bibinfo {author} {\bibfnamefont {L.~E.}\ \bibnamefont {Hueso}}, \ and\ \bibinfo {author} {\bibfnamefont {F.}~\bibnamefont {Casanova}},\ }\href {\doibase 10.1103/PhysRevLett.127.047202} {\bibfield  {journal} {\bibinfo  {journal} {Phys. Rev. Lett.}\ }\textbf {\bibinfo {volume} {127}},\ \bibinfo {pages} {047202} (\bibinfo {year} {2021})}\BibitemShut {NoStop}%
\bibitem [{\citenamefont {Han}\ \emph {et~al.}(2014)\citenamefont {Han}, \citenamefont {Kawakami}, \citenamefont {Gmitra},\ and\ \citenamefont {Fabian}}]{Han2014Oct}%
  \BibitemOpen
  \bibfield  {author} {\bibinfo {author} {\bibfnamefont {W.}~\bibnamefont {Han}}, \bibinfo {author} {\bibfnamefont {R.~K.}\ \bibnamefont {Kawakami}}, \bibinfo {author} {\bibfnamefont {M.}~\bibnamefont {Gmitra}}, \ and\ \bibinfo {author} {\bibfnamefont {J.}~\bibnamefont {Fabian}},\ }\href {\doibase 10.1038/nnano.2014.214} {\bibfield  {journal} {\bibinfo  {journal} {Nature Nanotechnology}\ }\textbf {\bibinfo {volume} {9}},\ \bibinfo {pages} {1748} (\bibinfo {year} {2014})}\BibitemShut {NoStop}%
\bibitem [{\citenamefont {Hu}\ and\ \citenamefont {Xiang}(2020)}]{Hu2020Dec}%
  \BibitemOpen
  \bibfield  {author} {\bibinfo {author} {\bibfnamefont {G.}~\bibnamefont {Hu}}\ and\ \bibinfo {author} {\bibfnamefont {B.}~\bibnamefont {Xiang}},\ }\href {\doibase 10.1186/s11671-020-03458-y} {\bibfield  {journal} {\bibinfo  {journal} {Nanoscale Research Letters}\ }\textbf {\bibinfo {volume} {15}},\ \bibinfo {pages} {1556} (\bibinfo {year} {2020})}\BibitemShut {NoStop}%
\bibitem [{\citenamefont {Gmitra}\ and\ \citenamefont {Fabian}(2017{\natexlab{a}})}]{JFabianOct2017}%
  \BibitemOpen
  \bibfield  {author} {\bibinfo {author} {\bibfnamefont {M.}~\bibnamefont {Gmitra}}\ and\ \bibinfo {author} {\bibfnamefont {J.}~\bibnamefont {Fabian}},\ }\href {\doibase 10.1103/PhysRevLett.119.146401} {\bibfield  {journal} {\bibinfo  {journal} {Phys. Rev. Lett.}\ }\textbf {\bibinfo {volume} {119}},\ \bibinfo {pages} {146401} (\bibinfo {year} {2017}{\natexlab{a}})}\BibitemShut {NoStop}%
\bibitem [{\citenamefont {Khoo}\ \emph {et~al.}(2017)\citenamefont {Khoo}, \citenamefont {Morpurgo},\ and\ \citenamefont {Levitov}}]{Levitov2017}%
  \BibitemOpen
  \bibfield  {author} {\bibinfo {author} {\bibfnamefont {J.~Y.}\ \bibnamefont {Khoo}}, \bibinfo {author} {\bibfnamefont {A.~F.}\ \bibnamefont {Morpurgo}}, \ and\ \bibinfo {author} {\bibfnamefont {L.}~\bibnamefont {Levitov}},\ }\href {\doibase 10.1021/acs.nanolett.7b03604} {\bibfield  {journal} {\bibinfo  {journal} {Nano Letters}\ }\textbf {\bibinfo {volume} {17}},\ \bibinfo {pages} {7003} (\bibinfo {year} {2017})}\BibitemShut {NoStop}%
\bibitem [{\citenamefont {McCann}\ and\ \citenamefont {Koshino}(2013)}]{McCann_2013}%
  \BibitemOpen
  \bibfield  {author} {\bibinfo {author} {\bibfnamefont {E.}~\bibnamefont {McCann}}\ and\ \bibinfo {author} {\bibfnamefont {M.}~\bibnamefont {Koshino}},\ }\href {\doibase 10.1088/0034-4885/76/5/056503} {\bibfield  {journal} {\bibinfo  {journal} {Reports on Progress in Physics}\ }\textbf {\bibinfo {volume} {76}},\ \bibinfo {pages} {056503} (\bibinfo {year} {2013})}\BibitemShut {NoStop}%
\bibitem [{\citenamefont {Slizovskiy}\ \emph {et~al.}(2019)\citenamefont {Slizovskiy}, \citenamefont {Garcia-Ruiz}, \citenamefont {Drummond},\ and\ \citenamefont {Falko}}]{Slizovskiy2019Dec}%
  \BibitemOpen
  \bibfield  {author} {\bibinfo {author} {\bibfnamefont {S.}~\bibnamefont {Slizovskiy}}, \bibinfo {author} {\bibfnamefont {A.}~\bibnamefont {Garcia-Ruiz}}, \bibinfo {author} {\bibfnamefont {N.}~\bibnamefont {Drummond}}, \ and\ \bibinfo {author} {\bibfnamefont {V.~I.}\ \bibnamefont {Falko}},\ }\href {https://arxiv.org/abs/1912.10067v1} {\bibfield  {journal} {\bibinfo  {journal} {arXiv}\ } (\bibinfo {year} {2019})},\ \Eprint {http://arxiv.org/abs/1912.10067} {1912.10067} \BibitemShut {NoStop}%
\bibitem [{\citenamefont {Young}\ and\ \citenamefont {Levitov}(2011)}]{Young2011}%
  \BibitemOpen
  \bibfield  {author} {\bibinfo {author} {\bibfnamefont {A.~F.}\ \bibnamefont {Young}}\ and\ \bibinfo {author} {\bibfnamefont {L.~S.}\ \bibnamefont {Levitov}},\ }\href {\doibase 10.1103/PhysRevB.84.085441} {\bibfield  {journal} {\bibinfo  {journal} {Phys. Rev. B}\ }\textbf {\bibinfo {volume} {84}},\ \bibinfo {pages} {085441} (\bibinfo {year} {2011})}\BibitemShut {NoStop}%
\bibitem [{\citenamefont {Wang}\ \emph {et~al.}(2019)\citenamefont {Wang}, \citenamefont {Che}, \citenamefont {Cao}, \citenamefont {Lyu}, \citenamefont {Watanabe}, \citenamefont {Taniguchi}, \citenamefont {Lau},\ and\ \citenamefont {Bockrath}}]{Wang2021}%
  \BibitemOpen
  \bibfield  {author} {\bibinfo {author} {\bibfnamefont {D.}~\bibnamefont {Wang}}, \bibinfo {author} {\bibfnamefont {S.}~\bibnamefont {Che}}, \bibinfo {author} {\bibfnamefont {G.}~\bibnamefont {Cao}}, \bibinfo {author} {\bibfnamefont {R.}~\bibnamefont {Lyu}}, \bibinfo {author} {\bibfnamefont {K.}~\bibnamefont {Watanabe}}, \bibinfo {author} {\bibfnamefont {T.}~\bibnamefont {Taniguchi}}, \bibinfo {author} {\bibfnamefont {C.~N.}\ \bibnamefont {Lau}}, \ and\ \bibinfo {author} {\bibfnamefont {M.}~\bibnamefont {Bockrath}},\ }\href {\doibase 10.1021/acs.nanolett.9b02445} {\bibfield  {journal} {\bibinfo  {journal} {Nano Letters}\ }\textbf {\bibinfo {volume} {19}},\ \bibinfo {pages} {7028} (\bibinfo {year} {2019})},\ \bibinfo {note} {pMID: 31525877}\BibitemShut {NoStop}%
\bibitem [{\citenamefont {Island}\ \emph {et~al.}(2019)\citenamefont {Island}, \citenamefont {Cui}, \citenamefont {Lewandowski}, \citenamefont {Khoo}, \citenamefont {Spanton}, \citenamefont {Zhou}, \citenamefont {Rhodes}, \citenamefont {Hone}, \citenamefont {Taniguchi}, \citenamefont {Watanabe}, \citenamefont {Levitov}, \citenamefont {Zaletel},\ and\ \citenamefont {Young}}]{Island2019}%
  \BibitemOpen
  \bibfield  {author} {\bibinfo {author} {\bibfnamefont {J.~O.}\ \bibnamefont {Island}}, \bibinfo {author} {\bibfnamefont {X.}~\bibnamefont {Cui}}, \bibinfo {author} {\bibfnamefont {C.}~\bibnamefont {Lewandowski}}, \bibinfo {author} {\bibfnamefont {J.~Y.}\ \bibnamefont {Khoo}}, \bibinfo {author} {\bibfnamefont {E.~M.}\ \bibnamefont {Spanton}}, \bibinfo {author} {\bibfnamefont {H.}~\bibnamefont {Zhou}}, \bibinfo {author} {\bibfnamefont {D.}~\bibnamefont {Rhodes}}, \bibinfo {author} {\bibfnamefont {J.~C.}\ \bibnamefont {Hone}}, \bibinfo {author} {\bibfnamefont {T.}~\bibnamefont {Taniguchi}}, \bibinfo {author} {\bibfnamefont {K.}~\bibnamefont {Watanabe}}, \bibinfo {author} {\bibfnamefont {L.~S.}\ \bibnamefont {Levitov}}, \bibinfo {author} {\bibfnamefont {M.~P.}\ \bibnamefont {Zaletel}}, \ and\ \bibinfo {author} {\bibfnamefont {A.~F.}\ \bibnamefont {Young}},\ }\href {\doibase 10.1038/s41586-019-1304-2} {\bibfield  {journal} {\bibinfo  {journal} {Nature}\ }\textbf {\bibinfo {volume} {571}},\ \bibinfo {pages} {85}
  (\bibinfo {year} {2019})}\BibitemShut {NoStop}%
\bibitem [{\citenamefont {Masseroni}\ \emph {et~al.}(2024)\citenamefont {Masseroni}, \citenamefont {Gull}, \citenamefont {Panigrahi}, \citenamefont {Jacobsen}, \citenamefont {Fischer}, \citenamefont {Tong}, \citenamefont {Gerber}, \citenamefont {Niese}, \citenamefont {Taniguchi}, \citenamefont {Watanabe}, \citenamefont {Levitov}, \citenamefont {Ihn}, \citenamefont {Ensslin},\ and\ \citenamefont {Duprez}}]{Masseroni2024Oct}%
  \BibitemOpen
  \bibfield  {author} {\bibinfo {author} {\bibfnamefont {M.}~\bibnamefont {Masseroni}}, \bibinfo {author} {\bibfnamefont {M.}~\bibnamefont {Gull}}, \bibinfo {author} {\bibfnamefont {A.}~\bibnamefont {Panigrahi}}, \bibinfo {author} {\bibfnamefont {N.}~\bibnamefont {Jacobsen}}, \bibinfo {author} {\bibfnamefont {F.}~\bibnamefont {Fischer}}, \bibinfo {author} {\bibfnamefont {C.}~\bibnamefont {Tong}}, \bibinfo {author} {\bibfnamefont {J.~D.}\ \bibnamefont {Gerber}}, \bibinfo {author} {\bibfnamefont {M.}~\bibnamefont {Niese}}, \bibinfo {author} {\bibfnamefont {T.}~\bibnamefont {Taniguchi}}, \bibinfo {author} {\bibfnamefont {K.}~\bibnamefont {Watanabe}}, \bibinfo {author} {\bibfnamefont {L.}~\bibnamefont {Levitov}}, \bibinfo {author} {\bibfnamefont {T.}~\bibnamefont {Ihn}}, \bibinfo {author} {\bibfnamefont {K.}~\bibnamefont {Ensslin}}, \ and\ \bibinfo {author} {\bibfnamefont {H.}~\bibnamefont {Duprez}},\ }\href {\doibase 10.1038/s41467-024-53324-z} {\bibfield  {journal} {\bibinfo  {journal} {Nat. Commun.}\ }\textbf
  {\bibinfo {volume} {15}},\ \bibinfo {pages} {1} (\bibinfo {year} {2024})}\BibitemShut {NoStop}%
\bibitem [{\citenamefont {Seiler}\ \emph {et~al.}(2025)\citenamefont {Seiler}, \citenamefont {Zhumagulov}, \citenamefont {Zollner}, \citenamefont {Yoon}, \citenamefont {Urbaniak}, \citenamefont {Geisenhof}, \citenamefont {Watanabe}, \citenamefont {Taniguchi}, \citenamefont {Fabian}, \citenamefont {Zhang},\ and\ \citenamefont {Weitz}}]{Seiler2024Mar}%
  \BibitemOpen
  \bibfield  {author} {\bibinfo {author} {\bibfnamefont {A.~M.}\ \bibnamefont {Seiler}}, \bibinfo {author} {\bibfnamefont {{\relax Ya}.}~\bibnamefont {Zhumagulov}}, \bibinfo {author} {\bibfnamefont {K.}~\bibnamefont {Zollner}}, \bibinfo {author} {\bibfnamefont {C.}~\bibnamefont {Yoon}}, \bibinfo {author} {\bibfnamefont {D.}~\bibnamefont {Urbaniak}}, \bibinfo {author} {\bibfnamefont {F.~R.}\ \bibnamefont {Geisenhof}}, \bibinfo {author} {\bibfnamefont {K.}~\bibnamefont {Watanabe}}, \bibinfo {author} {\bibfnamefont {T.}~\bibnamefont {Taniguchi}}, \bibinfo {author} {\bibfnamefont {J.}~\bibnamefont {Fabian}}, \bibinfo {author} {\bibfnamefont {F.}~\bibnamefont {Zhang}}, \ and\ \bibinfo {author} {\bibfnamefont {R.~T.}\ \bibnamefont {Weitz}},\ }\href {\doibase 10.1088/2053-1583/add74a} {\bibfield  {journal} {\bibinfo  {journal} {2D Mater.}\ }\textbf {\bibinfo {volume} {12}},\ \bibinfo {pages} {035009} (\bibinfo {year} {2025})}\BibitemShut {NoStop}%
\bibitem [{\citenamefont {Tiwari}\ \emph {et~al.}(2021)\citenamefont {Tiwari}, \citenamefont {Srivastav},\ and\ \citenamefont {Bid}}]{Tiwari2021Mar}%
  \BibitemOpen
  \bibfield  {author} {\bibinfo {author} {\bibfnamefont {P.}~\bibnamefont {Tiwari}}, \bibinfo {author} {\bibfnamefont {S.~K.}\ \bibnamefont {Srivastav}}, \ and\ \bibinfo {author} {\bibfnamefont {A.}~\bibnamefont {Bid}},\ }\href {\doibase 10.1103/PhysRevLett.126.096801} {\bibfield  {journal} {\bibinfo  {journal} {Phys. Rev. Lett.}\ }\textbf {\bibinfo {volume} {126}},\ \bibinfo {pages} {096801} (\bibinfo {year} {2021})}\BibitemShut {NoStop}%
\bibitem [{\citenamefont {Gorbachev}\ \emph {et~al.}(2007)\citenamefont {Gorbachev}, \citenamefont {Tikhonenko}, \citenamefont {Mayorov}, \citenamefont {Horsell},\ and\ \citenamefont {Savchenko}}]{Gorbachev2007}%
  \BibitemOpen
  \bibfield  {author} {\bibinfo {author} {\bibfnamefont {R.~V.}\ \bibnamefont {Gorbachev}}, \bibinfo {author} {\bibfnamefont {F.~V.}\ \bibnamefont {Tikhonenko}}, \bibinfo {author} {\bibfnamefont {A.~S.}\ \bibnamefont {Mayorov}}, \bibinfo {author} {\bibfnamefont {D.~W.}\ \bibnamefont {Horsell}}, \ and\ \bibinfo {author} {\bibfnamefont {A.~K.}\ \bibnamefont {Savchenko}},\ }\href {\doibase 10.1103/PhysRevLett.98.176805} {\bibfield  {journal} {\bibinfo  {journal} {Phys. Rev. Lett.}\ }\textbf {\bibinfo {volume} {98}},\ \bibinfo {pages} {176805} (\bibinfo {year} {2007})}\BibitemShut {NoStop}%
\bibitem [{\citenamefont {Liao}\ \emph {et~al.}(2010)\citenamefont {Liao}, \citenamefont {Han}, \citenamefont {Wu},\ and\ \citenamefont {Yu}}]{Liao2010Oct}%
  \BibitemOpen
  \bibfield  {author} {\bibinfo {author} {\bibfnamefont {Z.-M.}\ \bibnamefont {Liao}}, \bibinfo {author} {\bibfnamefont {B.-H.}\ \bibnamefont {Han}}, \bibinfo {author} {\bibfnamefont {H.-C.}\ \bibnamefont {Wu}}, \ and\ \bibinfo {author} {\bibfnamefont {D.-P.}\ \bibnamefont {Yu}},\ }\href {\doibase 10.1063/1.3505310} {\bibfield  {journal} {\bibinfo  {journal} {Appl. Phys. Lett.}\ }\textbf {\bibinfo {volume} {97}},\ \bibinfo {pages} {163110} (\bibinfo {year} {2010})}\BibitemShut {NoStop}%
\bibitem [{\citenamefont {Engels}\ \emph {et~al.}(2014)\citenamefont {Engels}, \citenamefont {Terr{\ifmmode\acute{e}\else\'{e}\fi}s}, \citenamefont {Epping}, \citenamefont {Khodkov}, \citenamefont {Watanabe}, \citenamefont {Taniguchi}, \citenamefont {Beschoten},\ and\ \citenamefont {Stampfer}}]{Engels2014Sep}%
  \BibitemOpen
  \bibfield  {author} {\bibinfo {author} {\bibfnamefont {S.}~\bibnamefont {Engels}}, \bibinfo {author} {\bibfnamefont {B.}~\bibnamefont {Terr{\ifmmode\acute{e}\else\'{e}\fi}s}}, \bibinfo {author} {\bibfnamefont {A.}~\bibnamefont {Epping}}, \bibinfo {author} {\bibfnamefont {T.}~\bibnamefont {Khodkov}}, \bibinfo {author} {\bibfnamefont {K.}~\bibnamefont {Watanabe}}, \bibinfo {author} {\bibfnamefont {T.}~\bibnamefont {Taniguchi}}, \bibinfo {author} {\bibfnamefont {B.}~\bibnamefont {Beschoten}}, \ and\ \bibinfo {author} {\bibfnamefont {C.}~\bibnamefont {Stampfer}},\ }\href {\doibase 10.1103/PhysRevLett.113.126801} {\bibfield  {journal} {\bibinfo  {journal} {Phys. Rev. Lett.}\ }\textbf {\bibinfo {volume} {113}},\ \bibinfo {pages} {126801} (\bibinfo {year} {2014})}\BibitemShut {NoStop}%
\bibitem [{\citenamefont {Zhai}(2022)}]{Zhai2022May}%
  \BibitemOpen
  \bibfield  {author} {\bibinfo {author} {\bibfnamefont {X.}~\bibnamefont {Zhai}},\ }\href {\doibase 10.1103/PhysRevB.105.205429} {\bibfield  {journal} {\bibinfo  {journal} {Phys. Rev. B}\ }\textbf {\bibinfo {volume} {105}},\ \bibinfo {pages} {205429} (\bibinfo {year} {2022})}\BibitemShut {NoStop}%
\bibitem [{\citenamefont {Yang}\ \emph {et~al.}(2016)\citenamefont {Yang}, \citenamefont {Tu}, \citenamefont {Kim}, \citenamefont {Wu}, \citenamefont {Wang}, \citenamefont {Alicea}, \citenamefont {Wu}, \citenamefont {Bockrath},\ and\ \citenamefont {Shi}}]{Yang2016sep}%
  \BibitemOpen
  \bibfield  {author} {\bibinfo {author} {\bibfnamefont {B.}~\bibnamefont {Yang}}, \bibinfo {author} {\bibfnamefont {M.-F.}\ \bibnamefont {Tu}}, \bibinfo {author} {\bibfnamefont {J.}~\bibnamefont {Kim}}, \bibinfo {author} {\bibfnamefont {Y.}~\bibnamefont {Wu}}, \bibinfo {author} {\bibfnamefont {H.}~\bibnamefont {Wang}}, \bibinfo {author} {\bibfnamefont {J.}~\bibnamefont {Alicea}}, \bibinfo {author} {\bibfnamefont {R.}~\bibnamefont {Wu}}, \bibinfo {author} {\bibfnamefont {M.}~\bibnamefont {Bockrath}}, \ and\ \bibinfo {author} {\bibfnamefont {J.}~\bibnamefont {Shi}},\ }\href {\doibase 10.1088/2053-1583/3/3/031012} {\bibfield  {journal} {\bibinfo  {journal} {2D Materials}\ }\textbf {\bibinfo {volume} {3}},\ \bibinfo {pages} {031012} (\bibinfo {year} {2016})}\BibitemShut {NoStop}%
\bibitem [{\citenamefont {Afzal}\ \emph {et~al.}(2018)\citenamefont {Afzal}, \citenamefont {Khan}, \citenamefont {Nazir}, \citenamefont {Dastgeer}, \citenamefont {Aftab}, \citenamefont {Akhtar}, \citenamefont {Seo},\ and\ \citenamefont {Eom}}]{Afzal2018feb}%
  \BibitemOpen
  \bibfield  {author} {\bibinfo {author} {\bibfnamefont {A.~M.}\ \bibnamefont {Afzal}}, \bibinfo {author} {\bibfnamefont {M.~F.}\ \bibnamefont {Khan}}, \bibinfo {author} {\bibfnamefont {G.}~\bibnamefont {Nazir}}, \bibinfo {author} {\bibfnamefont {G.}~\bibnamefont {Dastgeer}}, \bibinfo {author} {\bibfnamefont {S.}~\bibnamefont {Aftab}}, \bibinfo {author} {\bibfnamefont {I.}~\bibnamefont {Akhtar}}, \bibinfo {author} {\bibfnamefont {Y.}~\bibnamefont {Seo}}, \ and\ \bibinfo {author} {\bibfnamefont {J.}~\bibnamefont {Eom}},\ }\href {\doibase 10.1038/s41598-018-21787-y} {\bibfield  {journal} {\bibinfo  {journal} {Scientific Reports}\ }\textbf {\bibinfo {volume} {8}},\ \bibinfo {pages} {3412} (\bibinfo {year} {2018})}\BibitemShut {NoStop}%
\bibitem [{\citenamefont {Jafarpisheh}\ \emph {et~al.}(2018)\citenamefont {Jafarpisheh}, \citenamefont {Cummings}, \citenamefont {Watanabe}, \citenamefont {Taniguchi}, \citenamefont {Beschoten},\ and\ \citenamefont {Stampfer}}]{Jafarpisheh2018Dec}%
  \BibitemOpen
  \bibfield  {author} {\bibinfo {author} {\bibfnamefont {S.}~\bibnamefont {Jafarpisheh}}, \bibinfo {author} {\bibfnamefont {A.~W.}\ \bibnamefont {Cummings}}, \bibinfo {author} {\bibfnamefont {K.}~\bibnamefont {Watanabe}}, \bibinfo {author} {\bibfnamefont {T.}~\bibnamefont {Taniguchi}}, \bibinfo {author} {\bibfnamefont {B.}~\bibnamefont {Beschoten}}, \ and\ \bibinfo {author} {\bibfnamefont {C.}~\bibnamefont {Stampfer}},\ }\href {\doibase 10.1103/PhysRevB.98.241402} {\bibfield  {journal} {\bibinfo  {journal} {Phys. Rev. B}\ }\textbf {\bibinfo {volume} {98}},\ \bibinfo {pages} {241402} (\bibinfo {year} {2018})}\BibitemShut {NoStop}%
\bibitem [{\citenamefont {Amann}\ \emph {et~al.}(2022)\citenamefont {Amann}, \citenamefont {V\"olkl}, \citenamefont {Rockinger}, \citenamefont {Kochan}, \citenamefont {Watanabe}, \citenamefont {Taniguchi}, \citenamefont {Fabian}, \citenamefont {Weiss},\ and\ \citenamefont {Eroms}}]{amann2021}%
  \BibitemOpen
  \bibfield  {author} {\bibinfo {author} {\bibfnamefont {J.}~\bibnamefont {Amann}}, \bibinfo {author} {\bibfnamefont {T.}~\bibnamefont {V\"olkl}}, \bibinfo {author} {\bibfnamefont {T.}~\bibnamefont {Rockinger}}, \bibinfo {author} {\bibfnamefont {D.}~\bibnamefont {Kochan}}, \bibinfo {author} {\bibfnamefont {K.}~\bibnamefont {Watanabe}}, \bibinfo {author} {\bibfnamefont {T.}~\bibnamefont {Taniguchi}}, \bibinfo {author} {\bibfnamefont {J.}~\bibnamefont {Fabian}}, \bibinfo {author} {\bibfnamefont {D.}~\bibnamefont {Weiss}}, \ and\ \bibinfo {author} {\bibfnamefont {J.}~\bibnamefont {Eroms}},\ }\href {\doibase 10.1103/PhysRevB.105.115425} {\bibfield  {journal} {\bibinfo  {journal} {Phys. Rev. B}\ }\textbf {\bibinfo {volume} {105}},\ \bibinfo {pages} {115425} (\bibinfo {year} {2022})}\BibitemShut {NoStop}%
\bibitem [{\citenamefont {Icking}\ \emph {et~al.}(2022)\citenamefont {Icking}, \citenamefont {Banszerus}, \citenamefont {Wörtche}, \citenamefont {Volmer}, \citenamefont {Schmidt}, \citenamefont {Steiner}, \citenamefont {Engels}, \citenamefont {Hesselmann}, \citenamefont {Goldsche}, \citenamefont {Watanabe}, \citenamefont {Taniguchi}, \citenamefont {Volk}, \citenamefont {Beschoten},\ and\ \citenamefont {Stampfer}}]{Icking2022Oct}%
  \BibitemOpen
  \bibfield  {author} {\bibinfo {author} {\bibfnamefont {E.}~\bibnamefont {Icking}}, \bibinfo {author} {\bibfnamefont {L.}~\bibnamefont {Banszerus}}, \bibinfo {author} {\bibfnamefont {F.}~\bibnamefont {Wörtche}}, \bibinfo {author} {\bibfnamefont {F.}~\bibnamefont {Volmer}}, \bibinfo {author} {\bibfnamefont {P.}~\bibnamefont {Schmidt}}, \bibinfo {author} {\bibfnamefont {C.}~\bibnamefont {Steiner}}, \bibinfo {author} {\bibfnamefont {S.}~\bibnamefont {Engels}}, \bibinfo {author} {\bibfnamefont {J.}~\bibnamefont {Hesselmann}}, \bibinfo {author} {\bibfnamefont {M.}~\bibnamefont {Goldsche}}, \bibinfo {author} {\bibfnamefont {K.}~\bibnamefont {Watanabe}}, \bibinfo {author} {\bibfnamefont {T.}~\bibnamefont {Taniguchi}}, \bibinfo {author} {\bibfnamefont {C.}~\bibnamefont {Volk}}, \bibinfo {author} {\bibfnamefont {B.}~\bibnamefont {Beschoten}}, \ and\ \bibinfo {author} {\bibfnamefont {C.}~\bibnamefont {Stampfer}},\ }\href {\doibase https://doi.org/10.1002/aelm.202200510} {\bibfield  {journal} {\bibinfo  {journal}
  {Advanced Electronic Materials}\ }\textbf {\bibinfo {volume} {8}},\ \bibinfo {pages} {2200510} (\bibinfo {year} {2022})}\BibitemShut {NoStop}%
\bibitem [{\citenamefont {Uslu}\ \emph {et~al.}(2024{\natexlab{a}})\citenamefont {Uslu}, \citenamefont {Ouaj}, \citenamefont {Tebbe}, \citenamefont {Nekrasov}, \citenamefont {Bertram}, \citenamefont {Sch{\ifmmode\ddot{u}\else\"{u}\fi}tte}, \citenamefont {Watanabe}, \citenamefont {Taniguchi}, \citenamefont {Beschoten}, \citenamefont {Waldecker},\ and\ \citenamefont {Stampfer}}]{Uslu2023Jun}%
  \BibitemOpen
  \bibfield  {author} {\bibinfo {author} {\bibfnamefont {J.-L.}\ \bibnamefont {Uslu}}, \bibinfo {author} {\bibfnamefont {T.}~\bibnamefont {Ouaj}}, \bibinfo {author} {\bibfnamefont {D.}~\bibnamefont {Tebbe}}, \bibinfo {author} {\bibfnamefont {A.}~\bibnamefont {Nekrasov}}, \bibinfo {author} {\bibfnamefont {J.~H.}\ \bibnamefont {Bertram}}, \bibinfo {author} {\bibfnamefont {M.}~\bibnamefont {Sch{\ifmmode\ddot{u}\else\"{u}\fi}tte}}, \bibinfo {author} {\bibfnamefont {K.}~\bibnamefont {Watanabe}}, \bibinfo {author} {\bibfnamefont {T.}~\bibnamefont {Taniguchi}}, \bibinfo {author} {\bibfnamefont {B.}~\bibnamefont {Beschoten}}, \bibinfo {author} {\bibfnamefont {L.}~\bibnamefont {Waldecker}}, \ and\ \bibinfo {author} {\bibfnamefont {C.}~\bibnamefont {Stampfer}},\ }\href {\doibase 10.1088/2632-2153/ad2287} {\bibfield  {journal} {\bibinfo  {journal} {Mach. Learn.: Sci. Technol.}\ }\textbf {\bibinfo {volume} {5}},\ \bibinfo {pages} {015027} (\bibinfo {year} {2024}{\natexlab{a}})}\BibitemShut {NoStop}%
\bibitem [{\citenamefont {Uslu}\ \emph {et~al.}(2024{\natexlab{b}})\citenamefont {Uslu}, \citenamefont {Nekrasov}, \citenamefont {Hermans}, \citenamefont {Beschoten}, \citenamefont {Leibe}, \citenamefont {Waldecker},\ and\ \citenamefont {Stampfer}}]{Uslu2024Dec}%
  \BibitemOpen
  \bibfield  {author} {\bibinfo {author} {\bibfnamefont {J.-L.}\ \bibnamefont {Uslu}}, \bibinfo {author} {\bibfnamefont {A.}~\bibnamefont {Nekrasov}}, \bibinfo {author} {\bibfnamefont {A.}~\bibnamefont {Hermans}}, \bibinfo {author} {\bibfnamefont {B.}~\bibnamefont {Beschoten}}, \bibinfo {author} {\bibfnamefont {B.}~\bibnamefont {Leibe}}, \bibinfo {author} {\bibfnamefont {L.}~\bibnamefont {Waldecker}}, \ and\ \bibinfo {author} {\bibfnamefont {C.}~\bibnamefont {Stampfer}},\ }\href {\doibase 10.48550/arXiv.2412.09333} {\bibfield  {journal} {\bibinfo  {journal} {arXiv}\ } (\bibinfo {year} {2024}{\natexlab{b}}),\ 10.48550/arXiv.2412.09333},\ \Eprint {http://arxiv.org/abs/2412.09333} {2412.09333} \BibitemShut {NoStop}%
\bibitem [{\citenamefont {Icking}\ \emph {et~al.}(2024)\citenamefont {Icking}, \citenamefont {Emmerich}, \citenamefont {Watanabe}, \citenamefont {Taniguchi}, \citenamefont {Beschoten}, \citenamefont {Lemme}, \citenamefont {Knoch},\ and\ \citenamefont {Stampfer}}]{Icking2024Sep}%
  \BibitemOpen
  \bibfield  {author} {\bibinfo {author} {\bibfnamefont {E.}~\bibnamefont {Icking}}, \bibinfo {author} {\bibfnamefont {D.}~\bibnamefont {Emmerich}}, \bibinfo {author} {\bibfnamefont {K.}~\bibnamefont {Watanabe}}, \bibinfo {author} {\bibfnamefont {T.}~\bibnamefont {Taniguchi}}, \bibinfo {author} {\bibfnamefont {B.}~\bibnamefont {Beschoten}}, \bibinfo {author} {\bibfnamefont {M.~C.}\ \bibnamefont {Lemme}}, \bibinfo {author} {\bibfnamefont {J.}~\bibnamefont {Knoch}}, \ and\ \bibinfo {author} {\bibfnamefont {C.}~\bibnamefont {Stampfer}},\ }\href {\doibase 10.1021/acs.nanolett.4c02463} {\bibfield  {journal} {\bibinfo  {journal} {Nano Lett.}\ }\textbf {\bibinfo {volume} {24}},\ \bibinfo {pages} {11454} (\bibinfo {year} {2024})}\BibitemShut {NoStop}%
\bibitem [{\citenamefont {Gmitra}\ and\ \citenamefont {Fabian}(2017{\natexlab{b}})}]{Gmitra2017Oct}%
  \BibitemOpen
  \bibfield  {author} {\bibinfo {author} {\bibfnamefont {M.}~\bibnamefont {Gmitra}}\ and\ \bibinfo {author} {\bibfnamefont {J.}~\bibnamefont {Fabian}},\ }\href {\doibase 10.1103/PhysRevLett.119.146401} {\bibfield  {journal} {\bibinfo  {journal} {Phys. Rev. Lett.}\ }\textbf {\bibinfo {volume} {119}},\ \bibinfo {pages} {146401} (\bibinfo {year} {2017}{\natexlab{b}})}\BibitemShut {NoStop}%
\bibitem [{\citenamefont {Kechedzhi}\ \emph {et~al.}(2007)\citenamefont {Kechedzhi}, \citenamefont {McCann}, \citenamefont {Fal'ko}, \citenamefont {Suzuura}, \citenamefont {Ando},\ and\ \citenamefont {Altshuler}}]{Kechedzhi2007Sep}%
  \BibitemOpen
  \bibfield  {author} {\bibinfo {author} {\bibfnamefont {K.}~\bibnamefont {Kechedzhi}}, \bibinfo {author} {\bibfnamefont {E.}~\bibnamefont {McCann}}, \bibinfo {author} {\bibfnamefont {V.~I.}\ \bibnamefont {Fal'ko}}, \bibinfo {author} {\bibfnamefont {H.}~\bibnamefont {Suzuura}}, \bibinfo {author} {\bibfnamefont {T.}~\bibnamefont {Ando}}, \ and\ \bibinfo {author} {\bibfnamefont {B.~L.}\ \bibnamefont {Altshuler}},\ }\href {\doibase 10.1140/epjst/e2007-00224-6} {\bibfield  {journal} {\bibinfo  {journal} {The European Physical Journal Special Topics}\ }\textbf {\bibinfo {volume} {148}},\ \bibinfo {pages} {1951} (\bibinfo {year} {2007})}\BibitemShut {NoStop}%
\bibitem [{\citenamefont {Lu}\ \emph {et~al.}(2011)\citenamefont {Lu}, \citenamefont {Shi},\ and\ \citenamefont {Shen}}]{Lu2011}%
  \BibitemOpen
  \bibfield  {author} {\bibinfo {author} {\bibfnamefont {H.-Z.}\ \bibnamefont {Lu}}, \bibinfo {author} {\bibfnamefont {J.}~\bibnamefont {Shi}}, \ and\ \bibinfo {author} {\bibfnamefont {S.-Q.}\ \bibnamefont {Shen}},\ }\href {\doibase 10.1103/PhysRevLett.107.076801} {\bibfield  {journal} {\bibinfo  {journal} {Phys. Rev. Lett.}\ }\textbf {\bibinfo {volume} {107}},\ \bibinfo {pages} {076801} (\bibinfo {year} {2011})}\BibitemShut {NoStop}%
\bibitem [{\citenamefont {Ili\ifmmode~\acute{c}\else \'{c}\fi{}}\ \emph {et~al.}(2019)\citenamefont {Ili\ifmmode~\acute{c}\else \'{c}\fi{}}, \citenamefont {Meyer},\ and\ \citenamefont {Houzet}}]{ilic2019may}%
  \BibitemOpen
  \bibfield  {author} {\bibinfo {author} {\bibfnamefont {S.}~\bibnamefont {Ili\ifmmode~\acute{c}\else \'{c}\fi{}}}, \bibinfo {author} {\bibfnamefont {J.~S.}\ \bibnamefont {Meyer}}, \ and\ \bibinfo {author} {\bibfnamefont {M.}~\bibnamefont {Houzet}},\ }\href {\doibase 10.1103/PhysRevB.99.205407} {\bibfield  {journal} {\bibinfo  {journal} {Phys. Rev. B}\ }\textbf {\bibinfo {volume} {99}},\ \bibinfo {pages} {205407} (\bibinfo {year} {2019})}\BibitemShut {NoStop}%
\bibitem [{\citenamefont {Bluhm}\ \emph {et~al.}(2019)\citenamefont {Bluhm}, \citenamefont {Brückel}, \citenamefont {Morgenstern}, \citenamefont {von Plessen},\ and\ \citenamefont {Stampfer}}]{BluhmBuch2019}%
  \BibitemOpen
  \bibfield  {author} {\bibinfo {author} {\bibfnamefont {H.}~\bibnamefont {Bluhm}}, \bibinfo {author} {\bibfnamefont {T.}~\bibnamefont {Brückel}}, \bibinfo {author} {\bibfnamefont {M.}~\bibnamefont {Morgenstern}}, \bibinfo {author} {\bibfnamefont {G.}~\bibnamefont {von Plessen}}, \ and\ \bibinfo {author} {\bibfnamefont {C.}~\bibnamefont {Stampfer}},\ }\href {\doibase doi:10.1515/9783110438321} {\emph {\bibinfo {title} {Electrons in Solids: Mesoscopics, Photonics, Quantum Computing, Correlations, Topology}}}\ (\bibinfo  {publisher} {De Gruyter},\ \bibinfo {year} {2019})\BibitemShut {NoStop}%
\bibitem [{\citenamefont {Zollner}\ \emph {et~al.}(2020)\citenamefont {Zollner}, \citenamefont {Gmitra},\ and\ \citenamefont {Fabian}}]{Zollner2020Nov}%
  \BibitemOpen
  \bibfield  {author} {\bibinfo {author} {\bibfnamefont {K.}~\bibnamefont {Zollner}}, \bibinfo {author} {\bibfnamefont {M.}~\bibnamefont {Gmitra}}, \ and\ \bibinfo {author} {\bibfnamefont {J.}~\bibnamefont {Fabian}},\ }\href {\doibase 10.1103/PhysRevLett.125.196402} {\bibfield  {journal} {\bibinfo  {journal} {Phys. Rev. Lett.}\ }\textbf {\bibinfo {volume} {125}},\ \bibinfo {pages} {196402} (\bibinfo {year} {2020})}\BibitemShut {NoStop}%
\bibitem [{\citenamefont {Zollner}\ and\ \citenamefont {Fabian}(2021)}]{Zollner2021Aug}%
  \BibitemOpen
  \bibfield  {author} {\bibinfo {author} {\bibfnamefont {K.}~\bibnamefont {Zollner}}\ and\ \bibinfo {author} {\bibfnamefont {J.}~\bibnamefont {Fabian}},\ }\href {\doibase 10.1103/PhysRevB.104.075126} {\bibfield  {journal} {\bibinfo  {journal} {Phys. Rev. B}\ }\textbf {\bibinfo {volume} {104}},\ \bibinfo {pages} {075126} (\bibinfo {year} {2021})}\BibitemShut {NoStop}%
\bibitem [{\citenamefont {Gerber}\ \emph {et~al.}(2025)\citenamefont {Gerber}, \citenamefont {Ersoy}, \citenamefont {Masseroni}, \citenamefont {Niese}, \citenamefont {Laumer}, \citenamefont {Denisov}, \citenamefont {Duprez}, \citenamefont {Huang}, \citenamefont {Adam}, \citenamefont {Ostertag}, \citenamefont {Tong}, \citenamefont {Taniguchi}, \citenamefont {Watanabe}, \citenamefont {Fal'ko}, \citenamefont {Ihn}, \citenamefont {Ensslin},\ and\ \citenamefont {Knothe}}]{Gerber2025Apr}%
  \BibitemOpen
  \bibfield  {author} {\bibinfo {author} {\bibfnamefont {J.~D.}\ \bibnamefont {Gerber}}, \bibinfo {author} {\bibfnamefont {E.}~\bibnamefont {Ersoy}}, \bibinfo {author} {\bibfnamefont {M.}~\bibnamefont {Masseroni}}, \bibinfo {author} {\bibfnamefont {M.}~\bibnamefont {Niese}}, \bibinfo {author} {\bibfnamefont {M.}~\bibnamefont {Laumer}}, \bibinfo {author} {\bibfnamefont {A.~O.}\ \bibnamefont {Denisov}}, \bibinfo {author} {\bibfnamefont {H.}~\bibnamefont {Duprez}}, \bibinfo {author} {\bibfnamefont {W.~W.}\ \bibnamefont {Huang}}, \bibinfo {author} {\bibfnamefont {C.}~\bibnamefont {Adam}}, \bibinfo {author} {\bibfnamefont {L.}~\bibnamefont {Ostertag}}, \bibinfo {author} {\bibfnamefont {C.}~\bibnamefont {Tong}}, \bibinfo {author} {\bibfnamefont {T.}~\bibnamefont {Taniguchi}}, \bibinfo {author} {\bibfnamefont {K.}~\bibnamefont {Watanabe}}, \bibinfo {author} {\bibfnamefont {V.~I.}\ \bibnamefont {Fal'ko}}, \bibinfo {author} {\bibfnamefont {T.}~\bibnamefont {Ihn}}, \bibinfo {author} {\bibfnamefont {K.}~\bibnamefont
  {Ensslin}}, \ and\ \bibinfo {author} {\bibfnamefont {A.}~\bibnamefont {Knothe}},\ }\href {\doibase 10.48550/arXiv.2504.05864} {\bibfield  {journal} {\bibinfo  {journal} {arXiv}\ } (\bibinfo {year} {2025}),\ 10.48550/arXiv.2504.05864},\ \Eprint {http://arxiv.org/abs/2504.05864} {2504.05864} \BibitemShut {NoStop}%
\bibitem [{\citenamefont {Dulisch}\ \emph {et~al.}(2025)\citenamefont {Dulisch}, \citenamefont {Emmerich}, \citenamefont {Icking}, \citenamefont {Hecker}, \citenamefont {M{\ifmmode\ddot{o}\else\"{o}\fi}ller}, \citenamefont {M{\ifmmode\ddot{u}\else\"{u}\fi}ller}, \citenamefont {Watanabe}, \citenamefont {Taniguchi}, \citenamefont {Volk},\ and\ \citenamefont {Stampfer}}]{Dulisch2025Apr}%
  \BibitemOpen
  \bibfield  {author} {\bibinfo {author} {\bibfnamefont {H.}~\bibnamefont {Dulisch}}, \bibinfo {author} {\bibfnamefont {D.}~\bibnamefont {Emmerich}}, \bibinfo {author} {\bibfnamefont {E.}~\bibnamefont {Icking}}, \bibinfo {author} {\bibfnamefont {K.}~\bibnamefont {Hecker}}, \bibinfo {author} {\bibfnamefont {S.}~\bibnamefont {M{\ifmmode\ddot{o}\else\"{o}\fi}ller}}, \bibinfo {author} {\bibfnamefont {L.}~\bibnamefont {M{\ifmmode\ddot{u}\else\"{u}\fi}ller}}, \bibinfo {author} {\bibfnamefont {K.}~\bibnamefont {Watanabe}}, \bibinfo {author} {\bibfnamefont {T.}~\bibnamefont {Taniguchi}}, \bibinfo {author} {\bibfnamefont {C.}~\bibnamefont {Volk}}, \ and\ \bibinfo {author} {\bibfnamefont {C.}~\bibnamefont {Stampfer}},\ }\href {\doibase 10.1021/acs.nanolett.5c02229} {\bibfield  {journal} {\bibinfo  {journal} {Nano Lett.}\ }\textbf {\bibinfo {volume} {2025}} (\bibinfo {year} {2025}),\ 10.1021/acs.nanolett.5c02229}\BibitemShut {NoStop}%
\bibitem [{\citenamefont {Albrecht}\ \emph {et~al.}(2017)\citenamefont {Albrecht}, \citenamefont {Moers},\ and\ \citenamefont {Hermanns}}]{Albrecht2017May}%
  \BibitemOpen
  \bibfield  {author} {\bibinfo {author} {\bibfnamefont {W.}~\bibnamefont {Albrecht}}, \bibinfo {author} {\bibfnamefont {J.}~\bibnamefont {Moers}}, \ and\ \bibinfo {author} {\bibfnamefont {B.}~\bibnamefont {Hermanns}},\ }\href {\doibase 10.17815/jlsrf-3-158} {\bibfield  {journal} {\bibinfo  {journal} {Journal of Large-Scale Research Facilities}\ }\textbf {\bibinfo {volume} {3}},\ \bibinfo {pages} {112} (\bibinfo {year} {2017})}\BibitemShut {NoStop}%
\end{thebibliography}
\end{document}

\title{Supporting Information - Weak localization as a probe of proximity-induced spin-split valence bands in BLG on \ch{WSe2}}
\author{E.~Icking}
\affiliation{JARA-FIT and 2nd Institute of Physics, RWTH Aachen University, 52074 Aachen, Germany,~EU}%
\affiliation{Peter Gr\"unberg Institute  (PGI-9), Forschungszentrum J\"ulich, 52425 J\"ulich,~Germany,~EU}

\author{F.~Wörtche}
\affiliation{JARA-FIT and 2nd Institute of Physics, RWTH Aachen University, 52074 Aachen, Germany,~EU}%

\author{A.W.~Cummings}
\affiliation{Institut Català de Nanociència i Nanotecnologia: Bellaterra, Barcelona, Spain,~EU}%

\author{A.~Wörtche}
\affiliation{JARA-FIT and 2nd Institute of Physics, RWTH Aachen University, 52074 Aachen, Germany,~EU}%

\author{K.~Watanabe}
\affiliation{Research Center for Functional Materials, 
National Institute for Materials Science, 1-1 Namiki, Tsukuba 305-0044, Japan
}
\author{T.~Taniguchi}
\affiliation{ 
International Center for Materials Nanoarchitectonics, 
National Institute for Materials Science,  1-1 Namiki, Tsukuba 305-0044, Japan
}%
\author{C.~Volk}
\affiliation{JARA-FIT and 2nd Institute of Physics, RWTH Aachen University, 52074 Aachen, Germany,~EU}%
\affiliation{Peter Gr\"unberg Institute  (PGI-9), Forschungszentrum J\"ulich, 52425 J\"ulich,~Germany,~EU}

\author{B. Beschoten}
\affiliation{JARA-FIT and 2nd Institute of Physics, RWTH Aachen University, 52074 Aachen, Germany,~EU}%
\author{C. Stampfer}
\affiliation{JARA-FIT and 2nd Institute of Physics, RWTH Aachen University, 52074 Aachen, Germany,~EU}%
\affiliation{Peter Gr\"unberg Institute  (PGI-9), Forschungszentrum J\"ulich, 52425 J\"ulich,~Germany,~EU}%

%\date{}
\maketitle

%\section{Introduction}
The first part of the Supporting Information presents Raman spectroscopy data (section~\ref{Raman}) used to determine the layer number of \ch{WSe2} and BLG. Measurements of the quantum Hall effect are used to extract the gate lever arm (section~\ref{QH}). The calculation for the mobility and mean-free path are shown in section~\ref{mobility}. In section~\ref{Phase} universal conductance fluctuations (UCFs) are used to extract the phase coherence length. Additional magnetoconductance measurements are shown in (section~\ref{Magnetoconductance}), followed by a description for a rough estimate of the spin splitting at the band edge (section~\ref{splitting}). The Hamiltonian used for the band structure can be found in section~\ref{Hamiltonian}, the calculation of the Berry phase in section~\ref{Berry}, and the model for the weak localization is shown in detail in section~\ref{model}. 

\section{Raman Spectrum}
\label{Raman}
\begin{figure}[h]
	 	\centering
   		 \includegraphics[width = 0.6\textwidth]{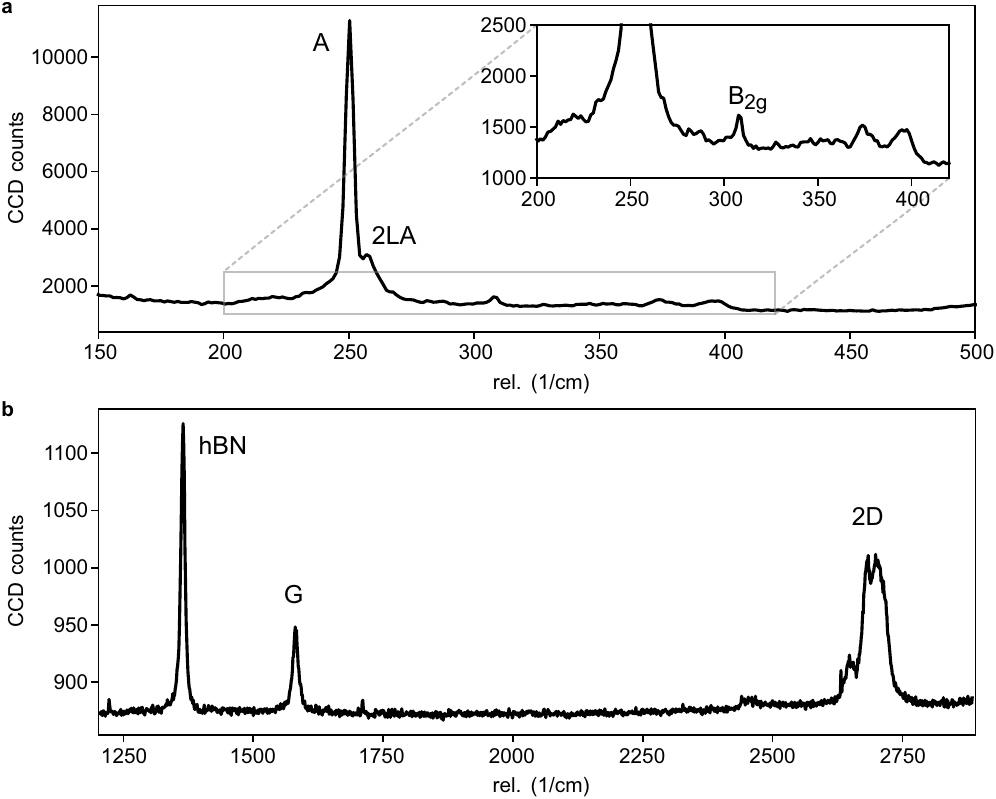}
	
		\caption{ Raman spectra of the hBN/BLG/\ch{WSe2}/hBN heterostructure. \textbf{a} Representative Raman spectrum of \ch{WSe2}. The B$_\text{2g}$ peak at 305 cm$^{-1}$ is an interlayer mode of \ch{WSe2}. \textbf{b} Representative Raman spectrum at higher wavenumber range, highlighting the characteristic   2D peak substructure, a distinct signature of high-quality BLG \cite{Schmitz2017Jun}.  } \label{fig:TMDraman}
\end{figure}

Raman spectroscopy is employed to confirm the layer stack and to extract the thickness of the \ch{WSe2} and BLG flakes in the device shown in the main text. In Fig.~\ref{fig:TMDraman}a we show the characteristic Raman peaks of \ch{WSe2} and in \ref{fig:TMDraman}b the respective Raman peaks for BLG). The B$_\text{2g}$ peak in the spectrum depicted in Fig.~\ref{fig:TMDraman}a at 305$\,$cm$^{-1}$ results from inter-layer vibrations in \ch{WSe2}~\cite{delCorro2014}. When comparing the shape and intensity of this peak to literature~\cite{delCorro2014}, we conclude that the flake is a bilayer. The 2D-peak in the spectrum Fig.~\ref{fig:TMDraman}b verifies that we used BLG in the heterostructure~\cite{Graf2007Feb, Neumann2015Sep,Schmitz2017Jun}.		
\FloatBarrier

\section{Gate lever arm determined from Quantum Hall effect measurement}
\label{QH}
Measuring the current in dependence of out-of-plane magnetic field and top gate voltage $V_\mathrm{tg}$ allows to extract the top gate lever arm~\cite{Zhao2010,Dauber2017,SonntagMay2018,Schmitz2020}:
\begin{equation}
    \alpha_\mathrm{tg}=\frac{\nu e B}{h V_\mathrm{tg}}+\text{constant}.
\end{equation}
Here, $\nu$ is the Landau filling factor, $B$ the applied magnetic field, $h$ the Planck constant. The extracted values are given in Table~\ref{tab:lever arm}.  
\begin{table}[]
    \centering
    \begin{tabular}{|c|c|c|c|}\hline
         $\alpha_\text{tg}(10^{11}$\,V$^{-1}$cm$^{-2})$ & $\beta$ & d$^\text{top}_\text{hBN}$ (nm) & d$^\text{bottom}_\text{hBN}$ (nm)\\\hline
         5.9 & 0.745 & 30 & 26\\\hline
    \end{tabular}
    \caption{List of top gate lever arm extracted from quantum Hall measurements, relative lever arm $\beta$, and thicknesses for top and bottom hBN layers from AFM measurements for the device presented in the main text.}
    \label{tab:lever arm}
\end{table}

		\begin{figure}[h]
	 	\centering
		 \includegraphics[width = 0.3\textwidth]{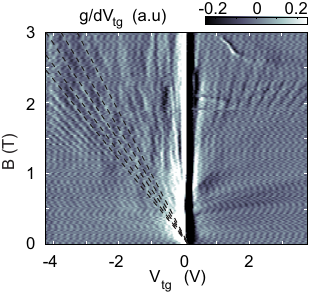}
		\caption[Landau-fan of the BLG-WSe2 device]{The derivative of the differential conductance $g=dI/dV$ as a function of top gate voltage $V_\mathrm{tg}$ and magnetic field $B$. From the slopes of different Landau-Level (see exemplarily the dashed lines), we can extract the gate lever arm $\alpha_\mathrm{tg}$.} \label{fig:QHE}
	\end{figure}

\section{Mobility and mean-free path}
\label{mobility}
\begin{figure}[hbt]
\centering
\includegraphics[draft=false,keepaspectratio=true,clip,width=0.3\linewidth]{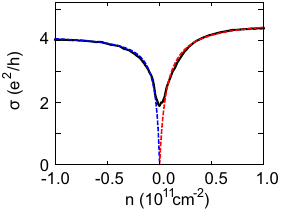}
\caption[Fig01]{ 
    Conductivity $\sigma$ as a function of the charge carrier density $n$. Red and blue curves represent fits of the conductivity for estimating the carrier mobility. 
}
\label{fig:mobility}
\end{figure}

Fig.~\ref{fig:mobility} shows a four-terminal conductivity trace $\sigma = \mathrm{d}I/\mathrm{d}V_\mathrm{sd} l/w$, measured as a function of charge carrier density $n$ at a back gate voltage $V_\mathrm{bg} = 0$\,V. The top gate region has a length $l=2$\,$\upmu$m and width $w=5$\,$\upmu$m. The charge carrier mobility $\mu_\text{e,h}$ for electrons and holes in the BLG is extracted by fitting the conductance using the relation $\sigma  = (\left(n e\mu_\text{e,h}\right)^{-1} + \rho_c)^{-1}$, where $\rho_c$ represents an additional series resistance originating from the ungated regions of the device. The top-gate-induced charge carrier density is given by $n = \alpha_\mathrm{tg} ( V_\mathrm{tg} - V_\mathrm{tg}^0)$. Notably, variations in $V_\mathrm{tg}$ affect not only $n$ but also the electric displacement field $D$, potentially introducing a small band gap.
However, within the applied voltage range, we expect electric displacement fields of only $D\approx \pm 20$\,mV/nm corresponding to a small band gap of $E_\text{g}\approx 1$\,meV, which is effectively negligible. 
From the conductance fit, we obtain mobilities of $\mu_{e} \approx 260,000$\,cm$^{2}$V$^{-1}$s$^{-1}$ and $\mu_{h} \approx 180,000$\,cm$^{2}$V$^{-1}$s$^{-1}$ for electrons and holes, respectively. 
The mean free path $l_\mathrm{m}$ of the charge carriers is calculated via  
    \begin{align}
        (l_\mathrm{m})_{e,h} = \frac{h}{2e} \times \sqrt{\frac{n}{\pi}} \times \mu_{e,h},
    \end{align}
with $h$ being the Planck constant.
For typical charge carrier densities used in the magneto-conductance measurements ranging from $n\approx\pm 1 \times 10^{11}$ to $\pm 4.5 \times 10^{11}$\,cm$^{-2}$, the estimated mean free paths are $l_{m,h} \approx 0.6-1.3\,\upmu$m for hole doping and $l_{m,e} \approx 1.0 - 2.0 \,\upmu$m for electron doping.

\section{Phase Coherence Length from Universal Conductance Oscillations}
\label{Phase}
\begin{figure}[h]

	 	\centering
	 	
   		 \includegraphics[width = 0.7\textwidth]{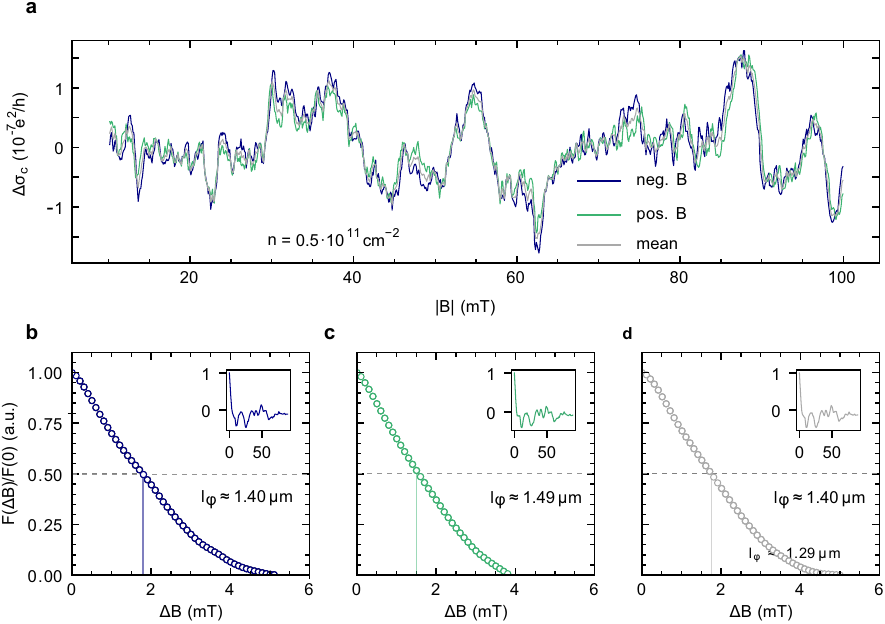}
		\caption{\textbf{a}  Universal conductance fluctuations (UCFs) and their autocorrelations at a gate-induced carrier density $n = 5 \times 10^{11}\,$cm$^{-2}$ and a displacement field D = -0.4 V/nm. The measurements were performed using a transconductance setup, and the obtained conductivity is corrected for a quadratic background, $\sigma_{BG}$. The conductance traces for negative (blue) and positive (green) magnetic fields are plotted along with their average (grey).  \textbf{b-d} show the corresponding normalized autocorrelation of the conductance traces. The phase coherence length $l_{\phi}$ is determined from the critical field $B_c$, defined as the magnetic field at which the normalized autocorrelation function drops to 0.5. To better visualize the position of $B_c$, the large panel shows a zoomed-in region of the autocorrelation, shown in its entirety in the inset.   }
		\label{UCF}
\end{figure}

\begin{figure}[h!]
	 	\centering
   		 \includegraphics[width = 0.5\textwidth]{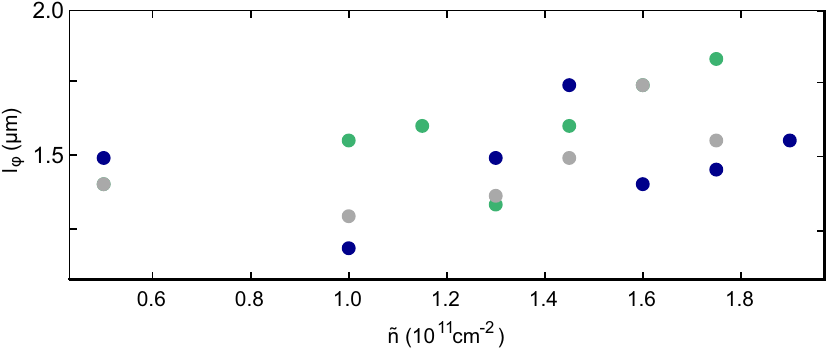}
		\caption{Phase coherence length determined from the  autocorrelation analysis of the universal conduction fluctuations, shown as a function of gate-induced charge carrier density at $D$ = -0.4\,V/nm.}
		\label{UCFn}
\end{figure}

As discussed in the main text, we use the universal conductance fluctuations (UCF) to extract the phase coherence length ($l_\phi$). UCFs, like WL and WAL, originate from quantum interference between all possible charge carriers can take. These effects become prominent when the phase coherence length is larger than the mean free path or for ballistic samples when $l_\phi$ is significantly larger than the characteristic size of the cavity, such that the paths can close or rejoin. When a perpendicular magnetic field is applied, the electron wave will experience a path-depending phase shift resulting in a fluctuation of the magneto-conductance measurements~\cite{BluhmBuch2019}. A typical length
scale of the fluctuations of the magnetoconductance can be obtained from the half maximum of the autocorrelation function $F(B,\Delta B)$ 
\begin{equation}
\label{autocorrelation}
    F(B,\Delta B)=\left\langle\delta\sigma(B,\Delta B)\delta\sigma(B)\right\rangle
\end{equation}
of the conductance fluctuation $\delta\sigma=\sigma(B)-\langle\sigma(B)\rangle$
with $\langle .. \rangle$ being the ensemble average~\cite{Lee1987UCFs,Zihlmann2018Feb}. As the variance of the fluctuation gives the maximum,
($F(B, \Delta B = 0)$), the correlation field is defined by $B_\mathrm{c} = 1/2 F(B; 0)$. Furthermore,  $B_\textrm{c}$ can be expressed by
\begin{equation}
    B_\mathrm{c}=const\cdot\frac{\Phi_0}{A_\phi}=\begin{cases}
const\cdot\frac{\Phi_0}{Wl_\phi} &\text{for $w$ $<l_\phi$}\\
const\cdot\frac{\Phi_0}{l^2_\phi} &\text{for $w$ $>l_\phi$}
\end{cases}
\end{equation}

with the flux quantum $\Phi_0 = h/e$ , the maximal enclosing phase coherent area $A_\phi$ and the width of the BLG channel. Given the top gate width of $5~\upmu$m, we operate in the regime $w > l_\phi$. 
Fig.~\ref{UCF}a shows the UCFs measured at $D = -0.4$~V/nm and ${n} = 1.65\times 10^{11}$~cm$^{-2}$. To isolate the fluctuation signal, a smooth background was removed using a second-order Savitzky–Golay filter applied across the full data range. To avoid contributions from WL/WAL effects, we restrict analysis to magnetic fields above 10~mT. Furthermore, we plot the curves for the positive (green curve) and negative magnetic field (blue curve) on top of each other and observe good agreement between them. To reduce the noise, the average of both is taken (depicted in gray). Applying the autocorrelation function given by Eq.~(\ref{autocorrelation}) to all UCF traces in Fig.~\ref{UCF}a results in a normalized autocorrelation as shown in Figs.~\ref{UCF}b to \ref{UCF}d. The main panel shows a zoom-in for $\Delta B\leq6$\,mT to highlight the position of $B_\mathrm{c}$, the trace over the full range of $\Delta B$ is depicted in the small inset. The extracted phase coherence lengths $l_\phi$ vary between $l_\phi=1.25$\,$\upmu$m and $l_\phi=1.75$\,$\upmu$m , see Fig.~\ref{UCFn}.

\newpage

\section{Magneto-transport Measurements}
\label{Magnetoconductance}

\begin{figure}[h!]
   
	 	\centering
   		 \includegraphics[width = 0.8\textwidth]{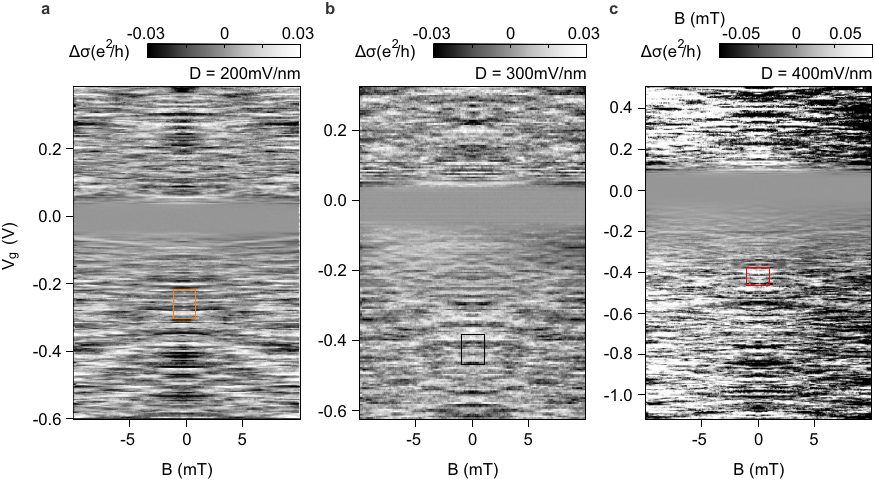}
		\caption{Magneto-conductance as a function of effective gate voltage $V_\text{g}$ and out-of-plane magnetic field at \textbf{a} $D=0.2$\,V/nm, \textbf{b} 0.3\,V/nm, and \textbf{c} 0.4\,V/nm.}
        \label{MC_posD}
	\end{figure}

\begin{figure}[h!]
   
	 	\centering
   		 \includegraphics[width = 0.4\textwidth]{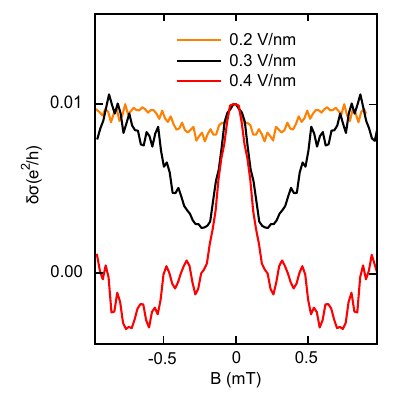}
		\caption{Averaged magneto-conductance trace as a function of out-of-plane magnetic field for $D=0.2$, 0.3, and 0.4\,V/nm. The $V_\text{g}$ interval, over which was averaged, is highlighted as colored boxes in Figure~\ref{MC_posD}.}
	\label{Fig:MC_posD-Wal}
    \end{figure}

In addition to the negative displacement fields discussed in the main text, we also show magneto-conductance measurements for positive displacement fields $D=0.2$, 0.3 and 0.4\,V/nm (see Figure~\ref{MC_posD}). For these positive values of $D$, we observe the WAL signal for hole doping (see Figure~\ref{Fig:MC_posD-Wal}), i.e., when a p-n junction is formed (see main text).

\newpage

\FloatBarrier

\section{Rough estimate of the spin splitting}
\label{splitting}
From the charger carrier interval $\Delta n$ we can estimate spin splitting $\Delta_\mathrm{SOC}=\hbar^2\pi \Delta n/(2m^*_\text{e})$, over which we observe WL. For a rough estimate we assume an effective mass $m^*_\text{e}\approx0.033\times m^*_\text{e}$. Furthermore, we assume $\Delta n \approx \left(\alpha_\text{tg}+\alpha_\text{bg}\right)\Delta V_\text{g}$ outside of the band gap.

\newpage
\section{Hamiltonian of BLG/TMD}\label{Hamiltonian}
In Fig.~4a of the manuscript, we show the low-energy bands of a BLG/\ch{WSe2} heterostructure for different ratios between the Rashba SOC strength $\lambda_\text{R}$ and the Valley-Zeeman SOC strength $\lambda_\text{VZ}$ at the K point ($\tau = 1$). The calculations are performed using the Hamiltonian of the BLG/TMD system, which is -- in the vicinity of the K or K' point of the Brillouin zone -- given by

\begin{eqnarray}
\hat{H} &=& \begin{bmatrix} \hat{H}_\text{bot} & \hat{H}_\text{int} \\ \hat{H}_\text{int}^\dag & \hat{H}_\text{top} + \hat{H}_\text{SOC} \end{bmatrix}, \nonumber ~~\\
\hat{H}_\text{top/bot} &=& \begin{bmatrix} \pm U &\xi \text{e}^{\text{i}\tau\theta} & 0 & 0 \\\xi \text{e}^{-\text{i}\tau\theta} & \pm U & 0 & 0 \\ 0 & 0 & \pm U &\xi \text{e}^{\text{i}\tau\theta} \\ 0 & 0 &\xi \text{e}^{-\text{i}\tau\theta} & \pm U \end{bmatrix}, \nonumber \\
\hat{H}_\text{int} &=& \begin{bmatrix} ~0 & t_\text{int} & 0 & 0 \\ ~0&0&0&0 \\ ~0&0&0&t_\text{int} \\ ~0&0&0&0 \end{bmatrix}, \nonumber ~~\\
\hat{H}_\text{SOC} &=& \begin{bmatrix} \tau\lvz & 0 & 0 & \text{i}\lr t_+ \\ 0 & \tau\lvz & \text{i}\lr t_- & 0 \\ 0 & -\text{i}\lr t_- & -\tau\lvz & 0 \\ -\text{i}\lr t_+ & 0 & 0 & -\tau\lvz \end{bmatrix},
\label{eq_ham}
\end{eqnarray}
with $\xi= \hbar v_F k \tau$ and $t_\pm=(\tau\pm 1)$.
$\hat{H}_\text{top/bot}$ is the single-layer graphene Hamiltonian describing the top (proximal) or bottom (distant) layer, where $v_F$ is the graphene Fermi velocity, $k$ and $\theta$ are the magnitude and direction of the electron momentum in the graphene plane, $\tau = \pm 1$ is the valley index for the K or K' valley, and $U$ is the interlayer potential arising from the out-of-plane electric displacement field. In the absence of SOC, the interlayer potential results in a band gap of $2U$. We assume an in-plane nearest-neighbor hopping of $t_0 = 2.7$ eV and a nearest-neighbor distance of $a_\text{cc} = 1.42$ \AA, giving $v = (3/2) a_\text{cc} t_0 / \hbar = 0.87 \times 10^6$ m/s. The interlayer coupling, given by $\hat{H}_\text{int}$, describes Bernal-stacked bilayer graphene with an interlayer hopping of $t_\text{int} = 0.4$ eV~\cite{Jung2014Jan}.\\
$\hat{H}_\text{SOC}$ describes the SOC induced in the graphene by proximity to the TMD, with $\lr$ the Rashba and $\lvz$ the valley-Zeeman SOC. Kane-Mele SOC may also be present, but \textit{ab initio} simulations show that this term tends to be much smaller than the others for this system \cite{Gmitra2016apr}, and we found it to have a negligible impact on our results, so here we neglect it. As indicated in Eq.~\eqref{eq_ham}, we assume that proximity to the TMD only affects the top (proximal) graphene layer. The Rashba SOC gives rise to a helical in-plane spin texture to the bands, while the valley-Zeeman term polarizes the spins out of the graphene plane, with opposite signs in opposite valleys \cite{Cummings2017Nov}. 

\section{Calculating the Berry phase}\label{Berry}
In general, the Berry phase is defined as the closed path integral of the differential phase of the wave function in parameter space~\cite{Resta1994}. Here we fix the absolute value of $k$ and consider a closed contour around the direction of momentum \cite{Lu2011},
\begin{equation}
\gamma_\text{B}(k) = -i\int\limits_0^{2\pi} \bra{\phi_k(\theta)} \frac{d}{d\theta} \ket{\phi_k(\theta)} d\theta,
\end{equation}
where $\ket{\phi_k(\theta)}$ is the eigenstate corresponding to the upper valence band at (integration radius) $k$ and direction $\theta$. Direct numerical integration of this expression can lead to instabilities, so we use the numerically gauge-invariant formulation \cite{Resta1994}
\begin{equation}
\gamma_\text{B}(k) = \Im \ln \prod_{n=0}^{N-1} \braket{\phi_k(\theta_n) | \phi_k(\theta_{n+1})},
\end{equation}
where $\theta_n = 2\pi n/N$ determine a discretized set of $N$ points
around the circular Fermi surface, and $\ket{\phi_k(\theta_i)}$ is the eigenstate corresponding to the upper valence band at momentum magnitude $k$ and direction $\theta_i$.

\section{Weak localization Model}
\label{model}
To model the amplitude of the measured WL signal as a function of charge carrier density we consider the following general expression for WL/WAL in bilayer graphene \cite{amann2021}
\begin{eqnarray}
\Delta\sigma(B) = &-&\frac{e^2}{2\pi h} \left[ F\left( \frac{\tb^{-1}}{\tphi^{-1}} \right) - F\left( \frac{\tb^{-1}}{\tphi^{-1} + 2\tasy^{-1}} \right) \right. \nonumber \\
&-&2F\left( \frac{\tb^{-1}}{\tphi^{-1} + \tasy^{-1} + \tsym^{-1}} \right) - F\left( \frac{\tb^{-1}}{\tphi^{-1} + 2\tiv^{-1}} \right) \nonumber \\
&-&2F\left( \frac{\tb^{-1}}{\tphi^{-1} + \tstar^{-1}} \right) + F\left( \frac{\tb^{-1}}{\tphi^{-1} + 2\tiv^{-1} + 2\tasy^{-1}} \right) \nonumber \\
&+&2F\left( \frac{\tb^{-1}}{\tphi^{-1} + \tstar^{-1} + 2\tasy^{-1}} \right) \nonumber \\
&+&2F\left( \frac{\tb^{-1}}{\tphi^{-1} + 2\tiv^{-1} + \tasy^{-1} + \tsym^{-1}} \right) \nonumber \\
&-&\left. 4F\left( \frac{\tb^{-1}}{\tphi^{-1} + \tstar^{-1} + \tasy^{-1} + \tsym^{-1}} \right) \right],
\label{eq_wal}
\end{eqnarray}
where $B$ is the magnetic field, $F(z) = \ln(z) + \psi(1/2 + 1/z)$, $\psi$ is the digamma function, $\tphi$ is the dephasing time, $\tb^{-1} = 4edB/\hbar$, $d$ is the diffusion coefficient, and $\tstar^{-1} = \tau_z^{-1} +  \tiv^{-1}$ is the elastic scattering rate with $\tau_z$ and $\tiv$ the intra- and intervalley scattering times respectively. The spin relaxation times $\tasy$ and $\tsym$ correspond to the SOC terms that break or conserve $z \rightarrow -z$ symmetry \cite{McCann2012apr}. In our model, $\tasy$ is thus associated with the Rasbha SOC, and $\tsym$ with the valley-Zeeman SOC.

To determine the various time scales in this expression, we assume that our device is sufficiently clean such that scattering and spin relaxation only occur at the top-gate-defined cavity and BLG edges. 
Thus, we let $\tau_z^{-1} = (l/v_\text{F})^{-1} + (w/v_\text{F})^{-1}$, $\tiv^{-1} = (\alpha l/v_\text{F})^{-1} + (w/v_\text{F})^{-1}$, and $d = \tstar v_\text{F}^2 / 2$, where $l = 2$ $\upmu$m and $w = 5$ $\upmu$m are the top gate defined channel length and width, and $v_\text{F}$ is the Fermi velocity of the hole states in the upper spin-split valence band. 
Here, we have assumed that the BLG edges are disordered enough to contribute equally to intra- and intervalley scattering. Meanwhile, the potential step at the cavity edges is somewhat softer and thus should contribute more weakly to intervalley scattering. This is reflected in the parameter $\alpha$, which we set to $\alpha=5$. Due to the helical nature of the in-plane spins induced by the Rashba SOC, the in-plane spin will be relaxed every time there is a change in carrier momentum; we thus let $\tasy = \tstar$. Similarly, the out-of-plane spin arising from the valley-Zeeman SOC has opposite signs in opposite valleys, giving $\tsym = \tiv$. 
We estimate $\tstar = \l_\phi^2/d\approx40$\,ps, based on the phase coherence length $l_\phi\approx1.5\upmu$m at $n\approx1.65\times10^{11}$~cm$^{-2}$.\\
The Fermi velocity $\vf$ is extracted from our band structure calculations. First, we get $v_\text{F}$ as a function of energy. We then numerically calculate the density of states and integrate it to get $n(E)$, such that we can translate $\vf(E)$ and $n(E)$ into $\vf(n)$. When only the upper valence band is occupied, the carrier velocity obtained from the model ranges from $10^3 - 10^5$ m/s, with higher velocity at higher carrier density.

\newpage